\documentclass[mathleft,fleqn]{an}
\usepackage{graphicx,color,natbib}
\usepackage{txfonts}
\usepackage{multirow}
\newcommand\cm{\,\rm cm}

\newcommand\g{\,\rm g}
\newcommand\erg{\,\rm erg}
\newcommand\K{\,\rm K}

\newcommand\Myr{\,\rm Myr}
\newcommand\Gyr{\,\rm Gyr}

\newcommand\kms{\,\rm km\,s^{-1}}
\newcommand\pc{\,\rm pc}

\newcommand\muG{\,\mu\rm G}
\newcommand\kpc{\,\rm kpc}

\newcommand\percc{\,\rm cm^{-3}}

\newcommand{\mn}[1]{\overline{#1}}

\newcommand\mB{\mn{\mathbf{B}}}

\begin{document}

\title{Dynamo saturation in direct simulations of the multi-phase
  turbulent interstellar medium} 

\titlerunning{dynamo saturation in ISM simulations}

\author{A. Bendre\inst{1}, O. Gressel\inst{2}, \and D. Elstner\inst{1}}

\institute{%
  Leibniz-Institut f\"ur Astrophysik Potsdam, 
  An der Sternwarte 16, 14482 Potsdam, Germany\\
  \email{abendre@aip.de} (AB), \email{delstner@aip.de} (DE)
  \and
  Niels Bohr International Academy, The Niels Bohr Institute, 
  Blegdamsvej 17, 2100 Copenhagen \O, Denmark\\
  \email{gressel@nbi.ku.dk} (OG)
}

\date{Received -- / Accepted --}

\abstract
{The ordered magnetic field observed via polarised synchrotron
  emission in nearby disc galaxies can be explained by a mean-field
  dynamo operating in the diffuse interstellar medium
  (ISM). Additionally, vertical-flux initial conditions are
  potentially able to influence this dynamo
  via the occurrence of the magnetorotational instability (MRI).
We aim to study the influence of various initial field
  configurations on the saturated state of the mean-field dynamo. This
  is motivated by the observation that different saturation behaviour
  was previously obtained for different supernova rates. 
 We perform direct numerical simulations (DNS) of three-dimensional 
  local boxes of the vertically
  stratified, turbulent interstellar medium, employing
  shearing-periodic boundary conditions horizontally. Unlike in our
  previous work, we also impose a vertical seed magnetic field. 
  We run the simulations until the growth of the magnetic energy 
  becomes negligible. We furthermore perform simulations of 
  equivalent 1D dynamo models, with an algebraic quenching
  mechanism for the dynamo coefficients.
We compare the saturation of the magnetic field in the DNS with the algebraic quenching of a mean-field dynamo. 
The final magnetic field strength found in the direct simulation is in
  excellent agreement with a quenched $\alpha\Omega$~dynamo. For supernova rates representative of the Milky Way,
  field losses via a Galactic wind are likely responsible for saturation.
 We conclude that the relative strength of the turbulent and regular
  magnetic fields in spiral galaxies may depend on the galaxy's star
  formation rate.  We propose that a mean field approach with
  algebraic quenching may serve as a simple sub-grid scale model for
  galaxy evolution simulations including a prescribed feedback from
  magnetic fields.}


\maketitle


\section{Introduction}

Magnetic fields in galaxies show coherent structures over kilo-parsec 
scales \citep{2013pss5.book..641B, 2010ASPC..438..197F}. The magnetic 
energy is comparable to the turbulent energy of the diffuse 
interstellar medium (ISM). The action of a mean-field dynamo is 
likely the main process for large-scale field generation 
\citep{1971ApJ...163..255P,1996ARA&A..34..155B,2005LNP...664..113S,2005PhR...417....1B}. 
Although simple heuristic mean-field models can already explain 
many observational facts, as for instance, a dominant axisymmetric 
mode in most of the galaxies, the details of the amplification 
process in the multi-phase ISM have remained hidden.  By means of 
direct numerical  simulations of the turbulent ISM, 
\citet{2008A&A...486L..35G} demonstrated the possibility of 
efficient field amplification through the dynamo effect on fast 
timescales of a few hundred million years.  But the non-linear 
saturation of the magnetic field is still an open issue.  Here 
we proceed with local box simulations of the turbulent ISM under 
the influence of rotation and shear, and we furthermore run these 
simulations until the fields become dynamically important. We are 
particularly interested in the behaviour of the dynamo closure 
coefficients in this regime, and we attempt to analyse their 
effect in the process of nonlinear saturation of the large-scale 
magnetic fields.

Energy injected by supernova (SN) explosions in combination with the
thermal instability \citep{1965ApJ...142..531F,2002ApJ...569L.127K}
gives rise to a turbulent multi-phase medium \citep[see,
  e.g.,][]{2005A&A...436..585D,2012AN....333..486B}.  We
furthermore look for an alternative origin of interstellar
turbulence, namely driving of turbulent fluctuations via the
magnetorotational instability (MRI). This has been proposed as a
source of turbulence in the outer parts of galaxies
\citep{1999ApJ...511..660S,2004A&A...423L..29D} but may be suppressed
by the turbulent diffusion from SNe
\citep{2010AN....331...34K,2013A&A...560A..93G}.  Exploring this
possibility, we hence extend the analysis performed by
\citet{2008A&A...486L..35G} to the case of a strong net-vertical flux
(NVF) for the seed fields. Neglecting the effect of energy input via
supernovae, \citet{2007ApJ...663..183P} have studied in detail the
multi-phase ISM with driving via the MRI alone. We accordingly avoid an
extensive discussion of the MRI in this paper, and instead focus on
the dynamo action produced by the SN driving. We currently also
neglect the contribution from cosmic rays
\citep{2006AN....327..469H,2009A&A...498..335H} -- we however plan to
address this question in the future.

Many other aspects like chemical evolution 
\citep{2014arXiv1412.2749W} or details of the energy transfer by 
radiation processes are beyond the scope of the paper.  The main 
goal here is to understand the amplification and the non-linear 
evolution of the magnetic field in the ISM. Compared to models 
with turbulence driven by an artificial forcing 
\citep[e.g.][]{2006MNRAS.370..415M}, our approach provides more 
realistic conditions.

This paper is structured as follows: In section~\ref{sec:model} we
describe the numerical methods and the physical model used to
represent a local patch of the turbulent, multi-phase ISM. General
results are presented in section~\ref{sec:res_general}. In
section~\ref{sec:dynamo} we describe the growth of the mean magnetic
field and its saturation. 
 We summarise our results and draw conclusions
in sections~\ref{sec:summary} and \ref{sec:concl}, respectively.


\section{Model Specifications and Methods}
\label{sec:model}

We solve the non-ideal magnetohydrodynamics (MHD) equations 
in a local co-rotating box of dimensions $0.8\times0.8\times4.2
\kpc^{3}$, where $x$ corresponds to the radial, $y$ to the azimuthal 
and $z$ to the vertical direction. A numerical resolution of $96
\times96 \times512$ grid cells is used  (grid size $\sim8.3\pc$). 
We use NIRVANA MHD fluid code by \citet{2004CoPhC.157..207Z} for 
our simulations. The set of equations we have used is described below 
(suppressing explicit factors of the constant  permeability
$\mu_0$, and with all symbols having their usual meanings):
\begin{eqnarray}
   \frac{\partial\rho}{\partial t} & = & 
      - \nabla\cdot\left(\rho \mathbf{u}\right) \,,
      \nonumber \\ 
      \frac{\partial\left(\rho\mathbf{u}\right)}{\partial t} & = & 
      - \nabla\cdot\left(\rho\mathbf{uu}
      + \mathrm{p}^{\star}
      - \mathbf{BB}\right) 
      - 2 \rho\ \Omega\ \hat{z}\times\mathbf{u}   \nonumber \\ & &
      + 2\rho\ \Omega^{2} q x\ \hat{x} 
      + \rho \mathrm{g}\ \hat{z} +\nabla\cdot\tau\,,
      \nonumber \\
      \frac{\partial e}{\partial t} & = &
      - \nabla\cdot\left((e+\mathrm{p}^{\star})\,\mathbf{u}-
      \left(\mathbf{u}\!\cdot\!\mathbf{B}\right)\mathbf{B}\right) 
      + 2 \rho\ \Omega^{2} q x\ \hat{x}\cdot\mathbf{u} \nonumber \\ & &
      + \rho \mathrm{g}\ \hat{z}\cdot\mathbf{u} 
      + \nabla\cdot\left[\tau \mathbf{u} 
      + \eta_{\rm m}\,\mathbf{B}\times\left(\nabla\times\mathbf{B}\right)\right] \nonumber \\ & &
      + \nabla\cdot(\kappa \nabla \mathrm{T})   
      - \rho^2 \Lambda\left(\mathrm{T}\right)
      + \Gamma_{\rm SN} + \rho\ \Gamma\left(z\right)\,,
      \nonumber \\
      \frac{\partial\mathbf{B}}{\partial t} & = & 
       \nabla\times\left(\mathbf{u}\times\mathbf{B} -\eta_{\rm m}
      \nabla\times\mathbf{B}\right)\,,
      \label{eq:mhd}
\end{eqnarray}
where the total pressure $\mathrm{p}^{\star}\equiv
\mathrm{p}+\mathbf{B}^2/2$, and assuming an adiabatic equation of
state, $\mathrm{p} = \left(\gamma-1\right)\ \epsilon$, with
$\gamma=5/3$, as appropriate for an ideal gas. Employing a total
energy formalism, the thermal energy density, $\epsilon$, is computed
from the total energy density (denoted by $e$) as,
\begin{eqnarray}
  \epsilon & = & e - \frac{\rho\mathbf{u}^2}{2} 
                   - \frac{\mathbf{B}^2}{2}\,,
\end{eqnarray}
and the viscous stress tensor is given by
\begin{eqnarray}
  \tau & = & \tilde{\nu}_{\rm m}\,\left(\nabla\mathbf{u}
            + (\nabla\mathbf{u})^{\top}
            - \frac{2}{3}\,(\nabla\cdot\mathbf{u})\right)\,,
\end{eqnarray}
where $\tilde{\nu}_{\rm m}$ represents the (molecular) 
dynamic viscosity coefficient, which is scaled with the density. 
We use a constant kinematic viscosity $\nu_{\rm m}=0.5 \times 
10^{25}\,{\rm cm^2s^{-1}}$, and a microscopic diffusivity of 
$\eta_{\rm m}= 2\times 10^{24}\,{\rm cm^2s^{-1}}$ such that the 
magnetic Prandtl number ${\rm Pm}\equiv \nu_{\rm m}/\eta_{\rm m}=2.5$ is 
somewhat larger than unity -- reflecting as good as feasible the 
fact that ${\rm Pm}\gg 1$ under typical ISM conditions.

The heat conduction term, $\nabla\cdot(\kappa \nabla \mathrm{T})$ with
a conductivity coefficient, $\kappa$, equivalent of a constant thermal
diffusivity, is included mainly for numerical reasons 
\citep[see discussion in][]{2009A&A...498..661G}. The various other
energy source and sink terms, as well as momentum source terms, are
described in the following sections.

\subsection{Supernova driving and thermodynamics} 
\label{sec:therm}
In our simulations, turbulence is driven by supernova explosions
(representing both type~I and type~II SNe), that are modelled via
localised injections of thermal energy. The vertical distributions of
the SN explosions are Gaussian with half widths of $325\pc$ and
$\sim90\pc$ for type~I/II SNe, respectively. The latter value is only
approximate, since, to avoid an artificial vertical dispersion of the
disc \citep{2008AN....329..619G}, we use the density profile
$\bar{\rho}(z)$ to compute a cumulative distribution function,
$\Phi_{\rm II}(z)$ via
\begin{equation}
  \Phi_{\rm II}(z) \equiv \frac{L_x\,L_y}{M}
  \int_{-L_z/2}^{z} \bar{\rho}(z')\,{\rm d}z'\,,
  \label{eq:vdistr}
\end{equation}
where $L_{x}$, $L_{y}$ and $L_{z}$ are the $x$, $y$ and $z$ dimensions
of the computational domain,respectively.  We then map an equally
distributed random number $r \in [0,1]$ via the inverse function,
$z=\Phi_{\rm II}^{-1}(r)$, to obtain a vertical distribution of
type~II SNe that follows the average mass density profile. The
reference Galactic supernova rates for type~I and II are $\sigma_{\rm I} =
4\Myr^{-1}\kpc^{-2}$  and $\sigma_{\rm II} = 30\Myr^{-1}\kpc ^{-2}$,
respectively, and associated energies are $10^{51}$ and $1.14
\times10^{51}\erg$. We furthermore mimic the effect of spatial
clustering for type~II SNe, which are known to occur in associations
of massive stars. To facilitate such a clustering, we resort to a very
crude prescription \citep{1999ApJ...514L..99K}: We first randomly
chose the vertical position of the explosion site according to
Eq.~(\ref{eq:vdistr}), and then determine random coordinates in the
horizontal plane. If the mass density at the chosen explosion site
exceeds the average mass density in that vertical slice, we add an
explosion, otherwise we chose a new random position in $(x,y)$ but
keeping the $z$~coordinate, until the criterion is met. We omit this
effect for type~I SN explosions.

A vertical profile of mass density with midplane value of $10^{-24} 
\g\cm^{-3}$ (equivalent to $1\cm^{-3}$) is used as an initial 
condition for the ISM density. The optically thin radiative 
cooling function is approximated by a piecewise power law as, 
$\Lambda(\mathrm{T})=\Lambda_{i}\ {\mathrm{T}}^{\beta_{i}}$, (
with ${\mathrm{T}}_{i}\leq \mathrm{T}\leq \mathrm{T}_{i+1}$).  
To capture the thermodynamics of the thermally unstable low-
temperature range, we implement the equilibrium pressure curve 
similar to \citet{2001ApJ...557L.121G} for $T<10^{5}\K$, which 
reproduces a well separated multi-phase ISM structure in rough 
agreement with the observed morphology. While choosing a 
piecewise-power-law approximation for the cooling function, we 
neglect the detailed non-equilibrium chemistry that will affect 
the cooling processes at high densities. This is reasonably 
justified because we are primarily interested in the dynamical 
aspects of the ISM at comparatively large scales and moderately 
high densities. Values of $\Lambda_{i}$ and $\beta_{i}$ are 
chosen in such a way that the temperature range between $141\K$ 
and $6102\K$ is prone to the thermal instability. Detailed 
discussion and corresponding fitting parameters ($\Lambda_{i}$ 
and $\beta_{i}$) for the cooling function are given in the section 
2.1 of \citet{2002ApJ...577..768S}. For the temperature 
$T>10^{-5}\K$, we use the cooling functions corresponding to 
the ISM in collisional ionisation equilibrium, similar to 
\citet{2005MNRAS.356..737S} (values of corresponding $\Lambda_{i}$ 
and $\beta_{i}$ are listed therein). \citet{2010PhDT.......244G} 
found that the presence of a thermally unstable branch in the 
cooling function only has a moderate influence on the large-scale 
dynamo.

Inverting the neutral $p(\rho)$ relation, this provides us with 
a natural choice for initial pressure profile, $p(z)$, 
simultaneously satisfying hydrostatic equilibrium and radiative 
energy balance (and leading to a slight temperature variation in 
the vertical direction).

\subsection{Differential rotation and gravity} 

In our local model, the background shear originating from the 
differential galactic rotation is expressed in terms of the shear 
parameter $q\equiv \mathrm{d}\ln \Omega/\mathrm{d}\ln R$, where 
$\mathrm{R}$ is the radius in a cylindrical coordinate system with 
its origin at the galactic centre. With this definition, $q=-1$ 
corresponds to a flat rotation curve.  We use initial conditions 
(midplane density and SN rate) corresponding to the solar orbit 
($\mathrm{R}\simeq 8.5\kpc$) for all our models. Because of the 
differentially rotating flow, we use shearing periodic boundary 
conditions \citep{2007CoPhC.176..652G} at the $x$ (radial) 
boundaries, whereas at the $y$ (azimuthal) and $z$ (vertical) 
boundaries, we use periodic and outflow boundary conditions, 
respectively. A vertical profile for acceleration due to gravity 
is chosen from \citet{1989ARA&A..27..555G}, that is
\begin{equation}
  \mathrm{g}(z) =
  \frac{-a_{1}\ z}{\sqrt{z^{2}
      -{z_{0}}^{2}}}-a_{2}\ z\,,
\end{equation}
with $a_{1}=1.42\times10^{-3}\kpc \Myr^{-2}$,
$a_{2}=5.49\times10^{-4}\Myr^{-2}$ and $z_{0}=180\pc$ -- with which we
tacitly neglect the effects of self gravity (which would presumably
only affect the ISM composition in the dense cold part).

\subsection{Description of studied models} 

In the first part of this analysis, we aim to study the effects of the
initial magnetic field configuration on the dynamo. We choose three
basic types of models, all with a supernova rate of $25\%$ in 
units of the average rate in the Milky Way $\sigma_{0}$, and where we 
keep the ratio of $\sigma_{\rm I}$ and $\sigma_{\rm II}$ fixed. The 
different models are characterised by
\begin{enumerate}
\itemsep4pt
\item[1.] weak vertical field ($0.001\muG$),\\ 
  with and without vertical flux (models `Q' and `QZ').
\item[2.] strong vertical field ($0.1\muG$),\\
  with and without vertical flux (`QS' and `QSZ').
\item[3.] weak azimuthal field ($0.001 \muG$),\\
  without vertical flux (model `AR').
\end{enumerate}
The description of the nomenclature (Q, QZ etc.) is given in 
Table~\ref{tab:models} as well.  It should be noted that the 
vertical flux of magnetic field in our box is conserved due 
to the periodic boundaries in the `${xy}$' plane. For 
reference, we also simulate a model with a zero net flux. A 
low SN rate together with a vertical flux magnetic field is 
chosen to include better the possible effects of MRI, which 
may be suppressed otherwise by a too strong turbulent 
diffusivity \citep[see e.g.][]{2013A&A...560A..93G}. To study 
the effect of SN driving on dynamo action and ISM composition 
in general, we simulate two models, with the same initial 
condition as Q, except we vary SN rates in this case. We call 
these models H and F, indicative of ``half'' (H) and ``full'' 
(F) supernova rate compared with the galactic SN rate 
$\sigma_{0}$. Assuming some kind of relation between column 
density and star formation rate, in reality, however, the SN 
rate does likely not vary without changing the mass density 
profile. Nevertheless it is worthwhile to establish how the 
dynamo process depends on the main driver for ISM turbulence. 
In the second part, we wish to study the effects of magnetic 
field and SN rate on the vertical structure.
\begin{table}[ht]
\caption{Model parameters for the simulations with vertical ($\rm
  B_v$) and in-plane ($\rm B_t$) initial fields. SN rates are
  in terms of the rate corresponding to the Milky Way,
  $\sigma_0 \equiv (\sigma_{I}^{-1} + \sigma_{II}^{-1})^{-1}$. The
  last column gives the physical time the models have been evolved
  for.}
\label{tab:models}
\centering
\begin{tabular}{lccccc}
\hline\\[-6pt]
    	&$\rm B_v$	&Flux		&$\rm B_t$	&SN Rate	&Time\\
	&$[\muG]$	&$[\muG\kpc^{-2}]$	&$[\muG]$	&$[\sigma_0]$	&[\Gyr]
\\[2pt]\hline\\[-6pt]
F	&0.001		&0.0064		&0		&1.00		&1.40\\
H	&0.001		&0.0064		&0		&0.50           &1.40\\
Q	&0.001		&0.0064		&0		&0.25           &2.08\\
QZ	&0.001		&0		&0		&0.25		&1.25\\
QS	&0.1		&0.064		&0		&0.25		&1.60\\
QSZ	&0.1		&0		&0		&0.25		&1.08\\
AR	&0		&0		&0.001		&0.25		&1.50
\\[2pt]\hline
\end{tabular}
\end{table}


\section{Results}
\label{sec:evol}

Starting from the initial model, which mostly consists of the warm
ionised and transition ISM phases ($5000\K$ to $10^{5.5}\K$), the
system evolves into a multi-phase turbulent state, that becomes
stationary in total thermal ($E_{\rm th}$) and kinetic energy ($E_{\rm kin}$)
within the first $20$ to $50\Myr$. The stationary value of turbulent kinetic
energy $E_{\rm kin}$ (not including the kinetic energy of the shear
flow) has an approximate dependence, $E_{\rm kin}
\sim\sigma^{0.8\pm0.04}$ on the SN rate $\sigma$. In contrast, $E_{\rm
  th}$ scales with $\sigma$ as $E_{\rm th}\sim\sigma^{0.5\pm0.05}$. 
The vertical profile of the average
density $\bar{\rho}(z)$ (that is, averaged over the ${xy}$
plane), also evolves to a steady state within the first $20$ to
$50\Myr$. With a quasi-stationary kinematic background state
established, we can now look into the temporal evolution of the
magnetic energy density.

\subsection{General evolution} 
\label{sec:res_general}

The initially Gaussian density profile (with an approximate 
scale height $\simeq 325\kpc$) evolves due to the influence 
of the developing turbulence, that adds a dynamical component 
to the vertical pressure support.  The resulting density 
profile shows a distinctive thin disc (with $\left|z\right| <
0.2\kpc$ which is unaffected by the SN rate), thick disc (within 
the range of $0.2 < \left|z\right| < 1\kpc$) and a halo 
component (above $1\kpc$). Accordingly, the best fit for this 
function is defined via a superposition of three exponential 
components, that is,

\begin{equation}
  \rho(z) \,\simeq\, \sum_{i=0}^{2}\; \rho_{i} \ \mathrm{exp}  
  \left(\frac{-|\,z\,|}{r_{i}}\right)\,.
  \label{eq:rho_fit}
\end{equation}

\smallskip\noindent Fitting coefficients $\rho_{i}$ and $r_{i}$ are listed in the
Table~\ref{tab:rho_scales} for the three runs Q, H, and F. Scale
heights in the halo (that is, $r_1$ and $r_2$) are found to increase
with the SN rate (with a dependence $\sim\sigma^{0.4}$), while $r_0$
does not appear to depend on $\sigma$. The equilibrium value of the
midplane density, $\rho_0$, of this component decreases slightly with
increasing SN rate. This is initial distribution does not change much 
under the influence of the magnetic field. 

\begin{table}[ht]
  \caption{Fitting coefficients for the $\bar{\rho}(z)$ profiles
    ($\rho_{i}$ and $r_{i}$).} \centering
  \begin{tabular}{lccc}
    \hline\\[-6pt]  
  
  Model			&Q		&H		&F		\\
  
  \\[2pt]\hline\\[-6pt]  
  $\rho_{0}$ $[\percc]$	&$0.6\pm0.1$	&$0.5\pm0.1$	&$0.3\pm0.1$	\\
  $r_{0}$ $[\kpc]$	&$0.11\pm0.04$	&$0.12\pm0.05$	&$0.12\pm0.05$	\\
  
  \hline\\[-6pt]
  $\rho_{1}$ $[\percc]$	&$0.1\pm0.03$	&$0.1\pm0.05$	&$0.1\pm0.05$		\\
  $r_{1}$ $[\kpc]$	&$0.25\pm0.04$	&$0.35\pm0.08$	&$0.42\pm0.1$	\\
  
  \hline\\[-6pt]
  $\rho_{2}$ $[\percc]$	&$0.01\pm0.002$	&$0.01\pm0.003$	&$0.01\pm0.005$	\\
  $r_{2}$ $[\kpc]$	&$0.7\pm0.3$	&$0.9\pm0.3$	&$1.2\pm0.5$	\\  
  \hline
  \end{tabular}\\[8pt]
  \label{tab:rho_scales}
\end{table}

\begin{figure}
  \center\includegraphics[width=0.9\columnwidth]{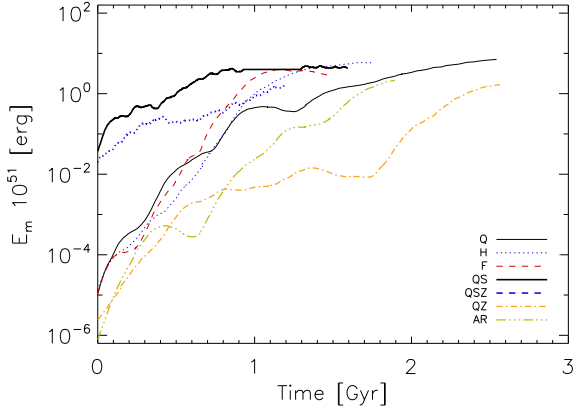}
  \caption{Evolution of the total magnetic energy for the various studied
    models.}
  \label{fig:tmag}
\end{figure}

After this initial phase of roughly $50\Myr$, the total magnetic
energy, $E_{\rm m}$, grows exponentially. This happens with almost the
same e-folding time, of about $100\Myr$, irrespective of SN rate for
several hundred million years. This is shown in Fig.~\ref{fig:tmag},
where we plot the time evolution of the magnetic energy. Quite
interestingly, the growth rate is slightly slower for the models with
higher initial field strengths (models QS, and QSZ), probably due to
non-linear quenching effects already becoming relevant.

The fast growth phase lasts until the total magnetic energies $E_{\rm m}$ are 6.6, 7.8 and $10.5 \times10^{50} {\erg}$ or 
the ratios to the kinetic energies $E_{\rm m}/E_{\rm kin}$ are $0.25$,
$0.14$ and $0.1$, in the three models with increasing supernova rate.
This state is reached after $1.2, 1.0$ and $0.8\Gyr$ for model Q, H,
and F, respectively. Afterwards, the growth rate continuously
decreases for models Q and H, but $E_{\rm m}$ keeps growing --
consistent with the derived algebraic quenching for the appropriate
mean-field model (cf. section \ref{sec:quench}).  For model F, the
growth stops after about $1\Gyr$. We shall refer to this phase as the
\emph{dynamical} state, in contrast to the initial \emph{kinematic}
growth phase.

\begin{table}
\centering
\caption{Temperature components of the ISM and average pressures at the mid plane $z=0$ for model Q}
\begin{tabular}{lcccc}
\hline\\[-6pt]
&T & $\rho$ & $\langle{P_{\rm th}} \rangle$& $\langle{P_{\rm kin}}\rangle$\\ 
& $[\K]$ & $[\cm^{-3}]$
& $[10^{-14}\ \mathrm{Pa}]$& $[10^{-14}\ \mathrm{Pa}]$
\\[2pt]\hline\\[-6pt]
Cold       & $  0-200 $          & $10$             & $2$   & $4$   \\ 
Cool       & $200-5000$          & $1$              & $5$   & $1.5$ \\ 
Warm     & $5000-10^{4.4}$     & $0.1$            & $6$   & $1.5$ \\ 
Trans     & $10^{4.4}-10^{5.5}$ & $5\times10^{-3}$ & $10$  & $15$  \\
Hot         & $>10^{5.5}$         & $5\times10^{-5}$ & $100$ & $50$  \\[2pt]
\hline
\end{tabular}
\label{tab:temp_comp}
\end{table}

\begin{table}
\caption{Composition of each ISM component at the end of the
    kinematic phase, i.e. at $1.2, 1.0$ and $0.9\Gyr$ for Q, H and 
    F respectively.  
  Energy fractions are calculated within the
  inner thin disc of $|z| < 0.5\kpc$, but Volume filling fractions (VFF) and mass filling fractions (MFF) are
  calculated for the entire box.}  \centering
\begin{tabular}{lcccccc}
\hline\\[-6pt]
                                &Cold   &Cool   &Warm   &Trans     &Hot    &Total  \\[2pt]
\hline\\[-6pt]
\multicolumn{7}{c}{model Q}\\[2pt]\hline\\[-6pt]
$E_{\rm m}/E_{\rm kin}$         &1.0    &1.2    &2.3    &0.6            &0.03   &1.6    \\
$E_{\rm m}/E_{\rm th}$          &2.0    &0.8    &1.0    &0.6            &0.01   &0.8    \\
$E_{\rm kin}/E_{\rm th}$        &1.7    &0.3    &0.4    &1.1            &0.4    &0.5    \\
VFF (\%)                        &0.02   &1.9    &20     &55             &23     &100    \\
MFF (\%)                        &3      &28     &62     &6.5            &0.25   &100    \\[2pt]
\hline\\[-6pt]
\multicolumn{7}{c}{model H}\\[2pt]\hline\\[-6pt]
$E_{\rm m}/E_{\rm kin}$         &0.5    &1.0    &1.2    &0.5            &0.03   &0.8    \\
$E_{\rm m}/E_{\rm th}$          &1.8    &0.6    &0.9    &0.5            &0.02   &0.6    \\
$E_{\rm kin}/E_{\rm th}$        &3.3    &0.6    &0.6    &1.1            &0.4    &0.7    \\
VFF (\%)                        &0.01   &1.7    &18     &50             &30     &100    \\
MFF (\%)                        &2      &26     &60     &11             &0.6    &100    \\[2pt]
\hline\\[-6pt]
\multicolumn{7}{c}{model F}\\[2pt]\hline\\[-6pt]
$E_{\rm m}/E_{\rm kin}$         &0.1    &0.5    &0.7    &0.35           &0.04   &0.2    \\
$E_{\rm m}/E_{\rm th}$          &0.7    &0.7    &0.8    &0.3            &0.02   &0.2    \\
$E_{\rm kin}/E_{\rm th}$        &8.3    &1.0    &1.0    &1.1            &0.4    &0.8    \\
VFF (\%)                        &0.006  &1.5    &14     &45             &40     &100    \\
MFF (\%)                        &1.6    &22     &55     &21             &1.4    &100    \\[2pt]
\hline
\end{tabular}
\label{tab:energy_content}
\end{table}

To illustrate the relative distribution of the different energy forms
with respect to the various ISM phases, we define five different
temperature ranges in Table~\ref{tab:temp_comp} and give the
distribution of magnetic, thermal and kinetic energy fractions within
these ISM phases in Table~\ref{tab:energy_content}, along with the
corresponding volume filling fractions (VFF) and mass filling
fractions (MFF).

\begin{figure*}
  \centering
  (a)
  \includegraphics[width=0.75\columnwidth]{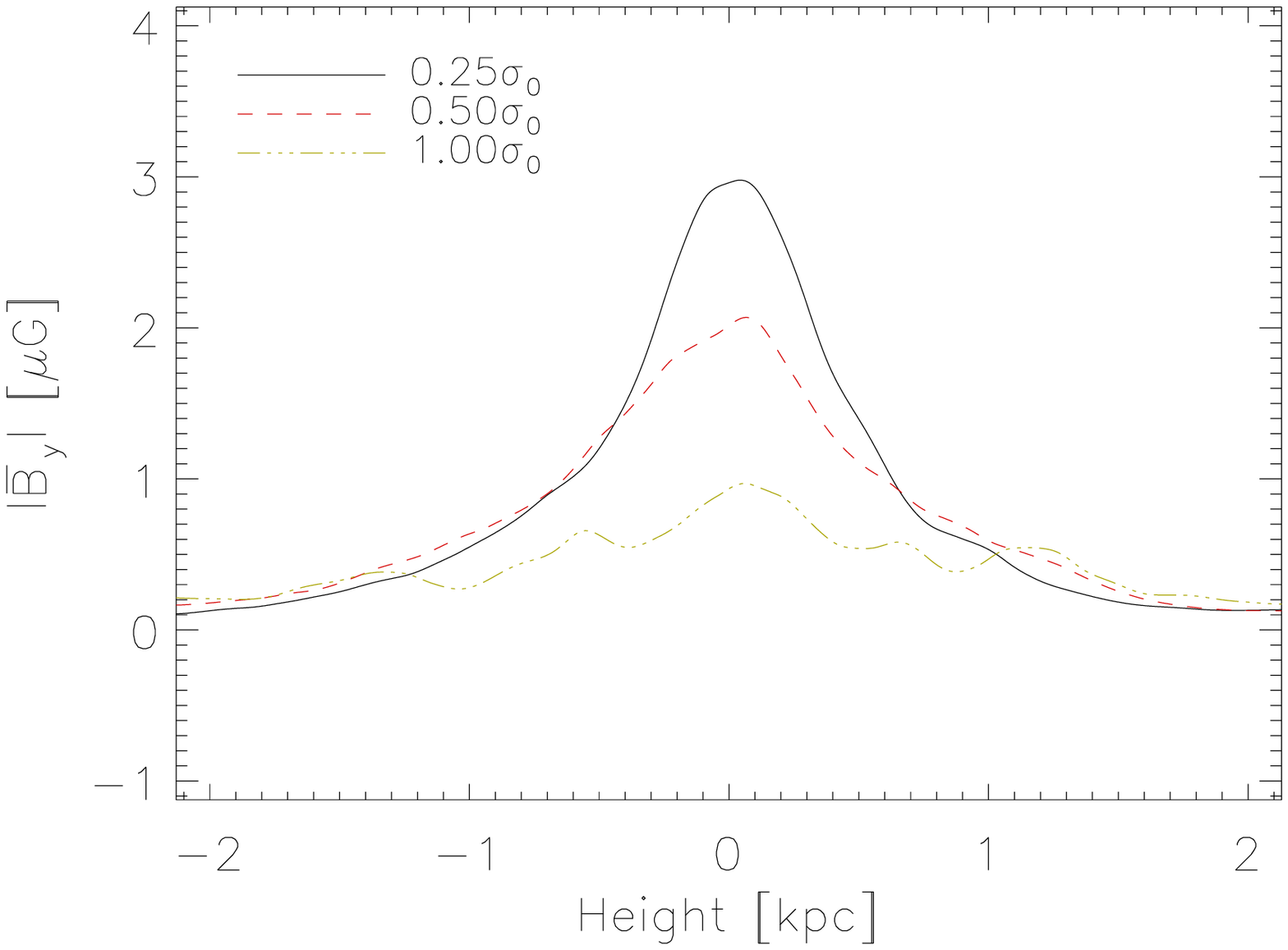}\hspace{20pt}
  (b)
  \includegraphics[width=0.75\columnwidth]{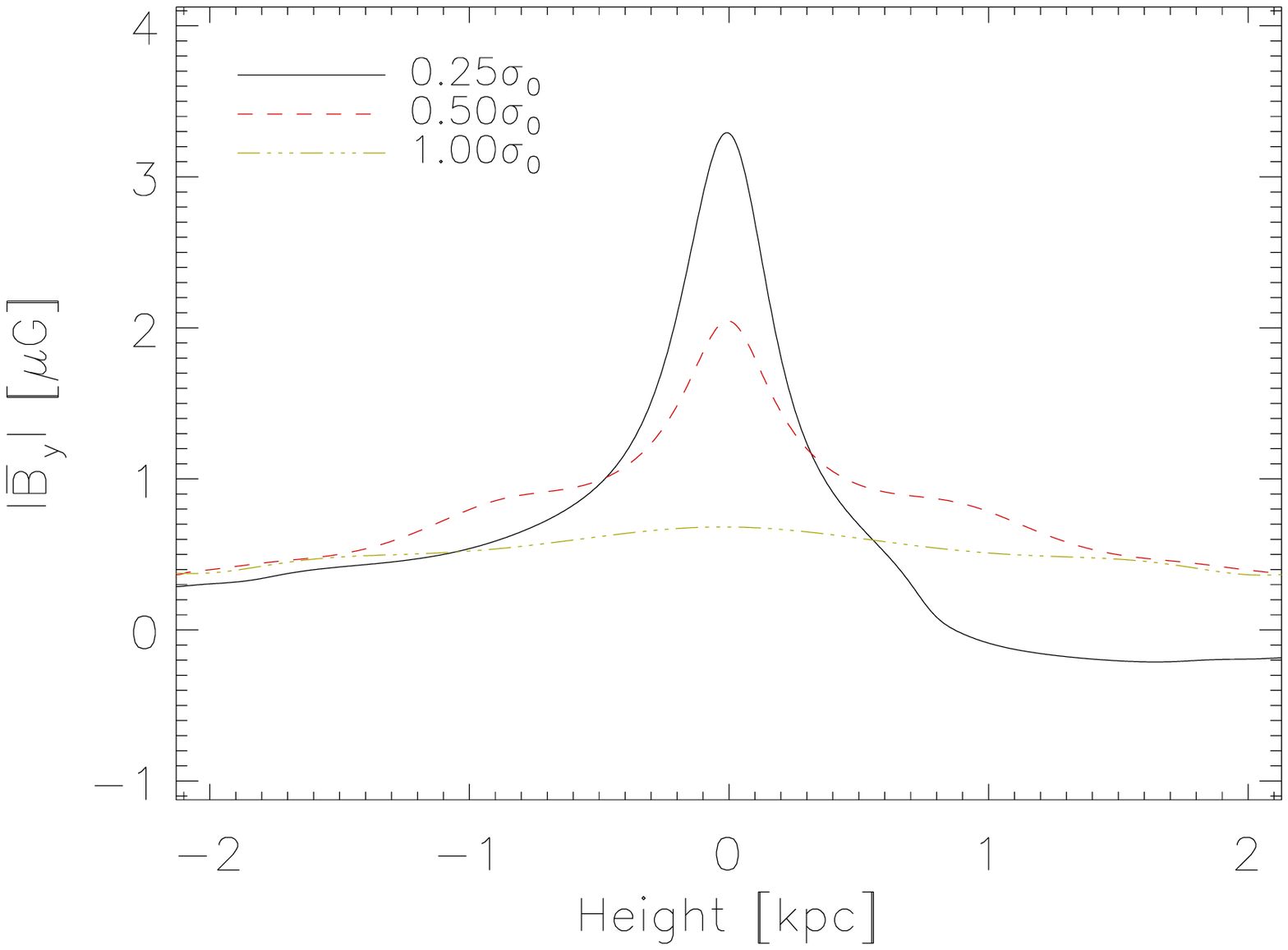}\\[10pt]
  (c)
  \includegraphics[width=0.75\columnwidth]{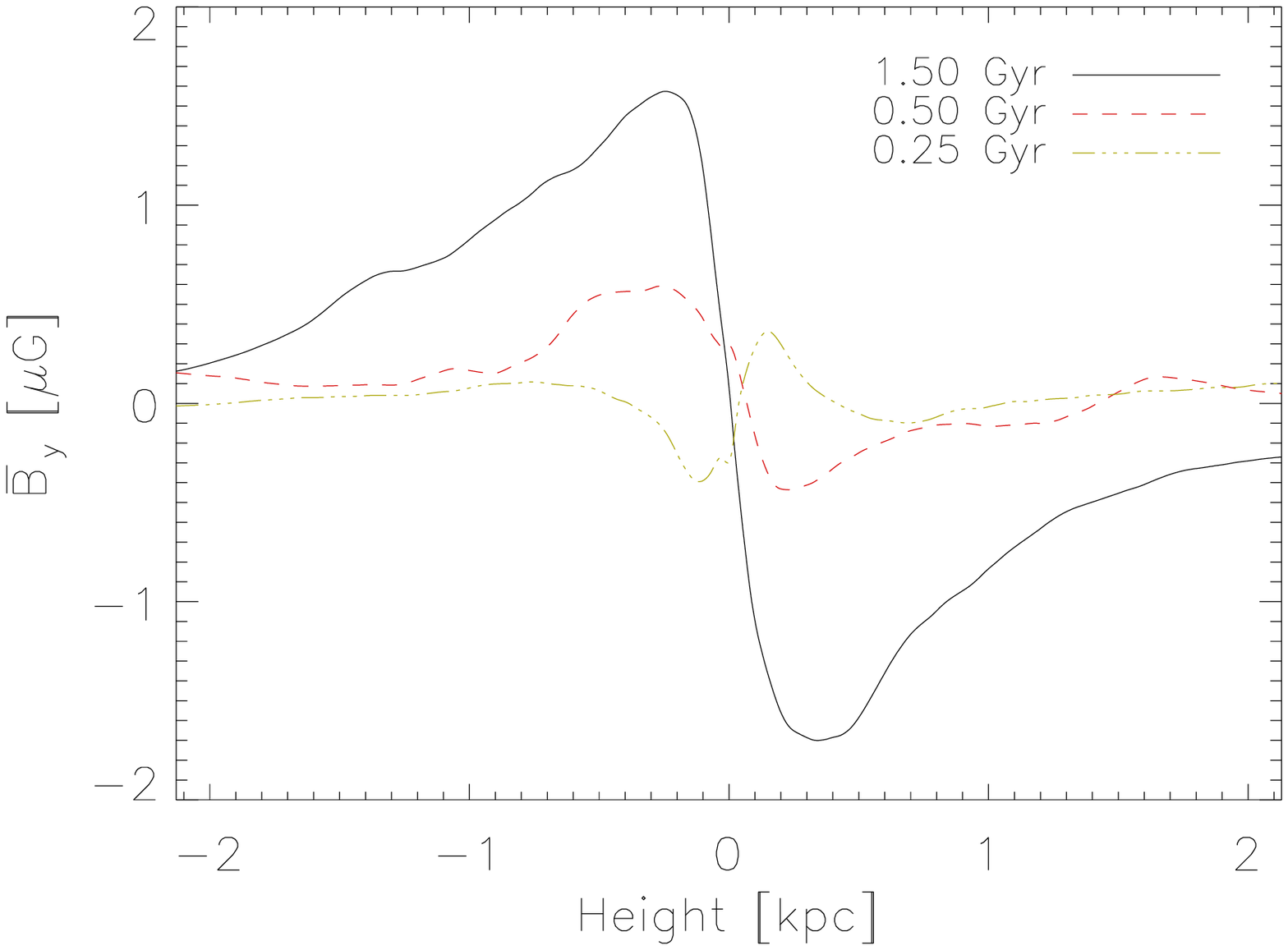}\hspace{20pt}
  (d)
  \includegraphics[width=0.75\columnwidth]{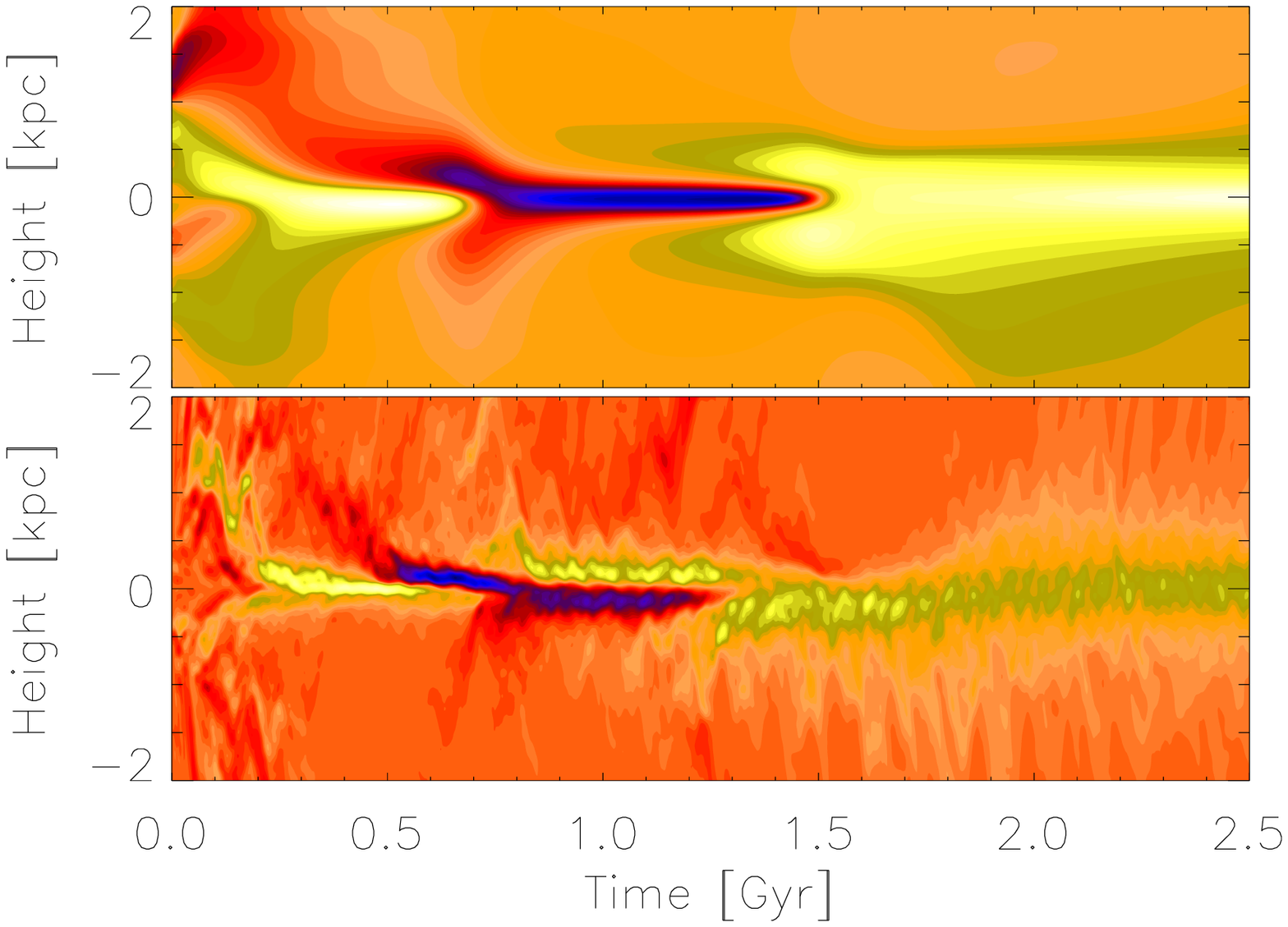}
  \caption{\emph{Upper-left}: Vertical profiles (from DNS) of the mean
    azimuthal magnetic field. Lines are for model Q (black; solid),
    for model H (red; dashed) and for model F (green; dot-dashed).
    \hspace{1ex}\emph{Upper-right}: Vertical profiles of the final 
    mean magnetic field from a 1D dynamo model (see section
    ~\ref{sec:mf_dynamo}).  \hspace{1ex}\emph{Lower-left}: Vertical 
    profiles of the mean magnetic field at different times, for the 
    strong net-flux model, QS, at low supernova rate. 
    \hspace{1ex}\emph{Lower-right}: Space-time-evolution of vertical 
    profiles of the azimuthal magnetic field for model Q, color 
    code indicates the strength of magnetic field normalised with the 
    square root of magnetic energy $E_m(t)$, which is to compensate 
    for the exponential growth in time and make the vertical mode 
    structure visible. The two panels compare the evolution seen in 
    the simple 1D dynamo model (upper panel, see section
    ~\ref{sec:mf_dynamo} for details) with the actual evolution in 
    the DNS (lower panel).}
  \label{fig:mean_fields}
\end{figure*}

\subsubsection{Mean magnetic field}
\label{sec:avg_b}

By averaging over horizontal planes, we split physical quantities 
like the total magnetic field, $\mathbf{B}$, into its mean part,

\begin{equation}
  \bar{\mathbf{B}}(z) = 
  \frac{1}{L_x L_y} \int\int{\mathbf{B}}\ {\rm d}x\,{\rm d}y\,,
\end{equation}
and fluctuating part, $\mathbf{b'} = \mathbf{B} - \bar{\mathbf{B}}(z)$.  
Due to the periodic and shearing periodic boundary conditions 
in the $y$ and $x$ directions respectively, the vertical ($z$) 
component of the magnetic flux remains unchanged throughout the 
evolution (subject to the solenoidal constraint). Moreover, we 
find that the $x$ component of $\bar{\mathbf{B}}$ remains 3 to 5 
times smaller than the $y$ component. For the purpose of brevity, we only present the azimuthal component in the plots. The 
vertical profile of $\bar{B}_y$ (and of $\bar{B}_x$) goes 
through several reversals and parity changes as it evolves in 
time, (see the lower right panel of Fig.~\ref{fig:mean_fields}). 
Finally, it achieves a steady-state S~mode (i.e, symmetric with 
respect to the midplane) in all models with weak (or zero) initial 
vertical flux (models Q, H, F, QZ, QSZ and AR).

The vertical shape found in DNS and a 1D mean-field model (cf. 
section~\ref{sec:mf_dynamo}) are shown in panels (a) and (b) of 
Fig.~\ref{fig:mean_fields} for models Q, H, and F, respectively. 
Vertical profiles at different times are shown for model QS in 
panel (c) of the same figure, along with the evolution of the 
mean azimuthal field for model Q, which is shown in panel (d). 
At late times, we obtain fits for these $\bar{B}_y(z)$ profiles, 
again applying a superposition of exponential functions with 
different scale heights, $h_{i}$, within the disc ($-0.8$ to 
$0.8\kpc$), and within the halo (above $0.8\kpc$).
\begin{equation}
  \bar{B}_y(z) \simeq \sum_{i=0}^{1} B_i
  \ \mathrm{exp}\left(-\frac{|\,z\,|}{h_{i}}\right),
\label{eq_mean_b_fit}
\end{equation}
values of the fitting coefficients ($B_{i}$ and $h_{i}$) are listed 
in Table~\ref{tab:b_scales}. We generally find the scale
heights to increase with increasing SN rate, reflecting the vertical
equilibration process with the additional kinetic pressure from the
SNe.

\begin{table}
  \caption{Fitting coefficients ($B_{i}$ and $h_{i}$) for the mean
    (azimuthal) magnetic field $\bar{B}_y(z)$.} \centering
  \begin{tabular}{lccccc}
    \hline\\[-6pt] 
    &$B_{0}$  		&$h_{0}$  	&$B_{1}$  	&$h_{1}$  	&$v_{\rm A}'(z\!=\!0)$ 	\\ 
    &$[\muG]$ 		&$[\kpc]$ 	&$[\muG]$	&$[\kpc]$ 	&$[\kms]$ 		\\[2pt]
    \hline\\[-6pt]
    $\!\!$Q &$1.8 \pm0.2$ 	&$0.5\pm0.1$ 	&$0.2\pm0.1$ 	&$1.2\pm0.4$ 	&$10\pm0.5$ 		\\ 
    $\!\!$H &$1.5 \pm0.2$ 	&$0.6\pm0.2$ 	&$0.3\pm0.1$ 	&$2.1\pm0.8$ 	&$14\pm0.5$ 		\\ 
    $\!\!$F &$0.7 \pm0.3$ 	&$0.8\pm0.2$ 	&$0.4\pm0.2$ 	&$3.6\pm1.4$ 	&$18\pm0.5$ 		\\ \hline
  \end{tabular}\\[8pt]

  \parbox[t]{\columnwidth}{\footnotesize \emph{Notes}: The scale
    heights for the rms $B(z)$ and mean azimuthal field, $\bar{B}_{y}(z)$ are
    nearly the same, consistent with a constant ratio of rms to mean field. 
    Midplane Alfv\'en velocities, $v_{\rm A}'(z=0)$,
    are stated at the end of kinematic phase, and roughly scale with
    the SN rate, $\sigma$, via a power law $v_{\rm A}' \sim
    \sigma^{0.4}$.}
  \label{tab:b_scales}
\end{table}

\noindent The midplane field strength, however, scales inversely with
respect to the SN rate (see the upper-left panel in
Fig.~\ref{fig:mean_fields}).  Moreover, flat regions in the energy
evolution curve, shown in Fig.~\ref{fig:tmag}, correspond to a flip in
the field direction, and these are also related to parity changes.
The model with high initial flux (QS which shows a strong `A' mode
initially) remains finally an `A' mode but with the same ratio of
$E_{\rm m}$ and $E_{\rm kin}$ as model Q. This is probably a
consequence of our boundary conditions, which conserve the vertical
magnetic flux. Initial growth times for QS and QSZ models are higher
compared with the weak flux models, which will be explained in the
next section.

\begin{figure}
  \center\includegraphics[width=0.9\columnwidth]{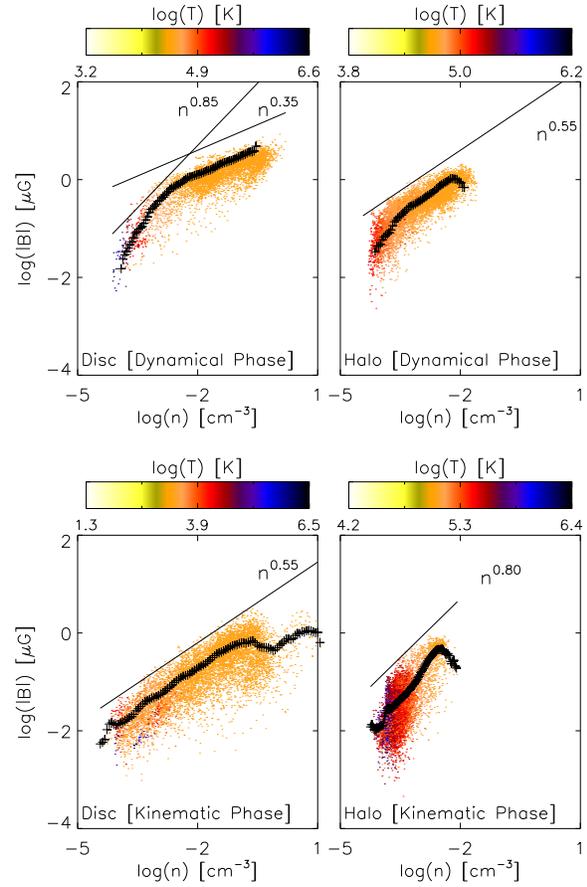}
  \caption{Typical distribution of the total magnetic field strength
    as a function of the mass density for model Q. The lower and upper
    rows compare the $B$-$\rho$~relation at early and late times,
    respectively, whereas the left and right columns compare the disc
    region $|z| < 0.5\kpc$ with the halo $|z| > 0.5\kpc$. The colour
    coding indicates the temperature in ${\rm K}$ as shown in the
    colour bar above each panel. `$+$' signs indicate the average
    value of total magnetic field for the corresponding density.
    Black straight lines in each panel show the best fit for the
    particular branch of the distribution, with power law exponents as
    shown.}
  \label{fig:B_rho_dist}
\end{figure}

\subsubsection{Fluctuating fields}

Root mean squared (rms) magnetic field profiles are defined by
\begin{equation}
  B(z) = \left[ \frac{1}{L_{x}L_{y}} \int
    \int{\left(\mathbf{B}\!\cdot\!\mathbf{B}\right)
    \ {\rm d}x\,{\rm d}y}\right]^{1/2} \,,
  \label{eq:rms_b_def}
\end{equation}
which are also best fitted with a function as the one in
Eq.~(\ref{eq_mean_b_fit}), and scale heights ($h_i$) which are
equivalent to the ones, that are listed in Table~\ref{tab:b_scales}
for the mean field.  We observe that the $B(z)$ profiles become wider
with SN rate, that is, the coefficients $h_{i}$ are directly
proportional to $\sigma$.  $B(z)$ have the same scale heights as those
of the mean fields, $\bar{B}_{y}(z)$, and midplane field strengths are
about $3\muG$ (at the end of kinematic phase) for all the models. By
virtue of field-line stretching and transverse compression,
amplification of $\mathbf{B}$ occurs in such a way that there exists a
statistical correlation between $B$ and $\rho$. To a first
approximation, this can be described as $B\propto\rho^{a}$, with a
certain exponent $a$, and where, for instance, $a=2/3$ would
correspond to amplification via compression in the two directions
perpendicular to the field line, and $a=1/2$ would correspond to
energy equipartition between magnetic and kinetic energy (assuming a
constant velocity dispersion).  From a scatter plot of $B$ versus
$\rho$, we find that $B$ scales with a different power law exponent
within the disc and in the halo. The coefficient is furthermore seen
to have a dependence on the field strength and SN rate. This is
represented in Fig.~\ref{fig:B_rho_dist}, in which we have plotted the
typical $B$-$\rho$ distribution during the kinematic as well as in the
dynamical phase for model Q.  First we consider only the disc midplane
region ($|z|<0.5\kpc$) during the kinematic phase. For this particular
regime, typical values of the exponent are $a\sim 0.5$. Later during
the dynamical phase, we obtain $a\sim 0.3$ for the dense ISM, that is
for $n> 0.01\percc$, whereas for the lighter ISM we obtain $a\sim 1$.
The flattening of the relation in the dense part may be explained by
the increasing dominance of the mean field over the fluctuating field.
For model F, the effect of a flattening of the relation is least
pronounced, and we find $a\sim0.4$ at late times (not shown) --
possibly because the quenching of the dynamo coefficients is smaller
compared with models Q and H (see Section~\ref{sec:mf_dynamo}).

In order to explain the comparatively steeper $B$-$\rho$ correlation
in the low-density ISM, we consider the scenario of flux
freezing. That is the regular field lines stretch out due to radially
expanding SN remnants (in the disc), and only negligible magnetic
fields remain inside the expanding shell (that is, in the low density
ISM). Now we consider the disc halo ($|z|>0.5\kpc$). For this region
$B$ scales as $\sim\rho^{0.8}$ during the initial kinematic phase
(similar to compressional amplification). Later  during the dynamical 
phase the magnetic field  density relation becomes again more flat with
$a\sim 0.5$.

\begin{figure}
  \center\includegraphics[width=0.9\columnwidth]{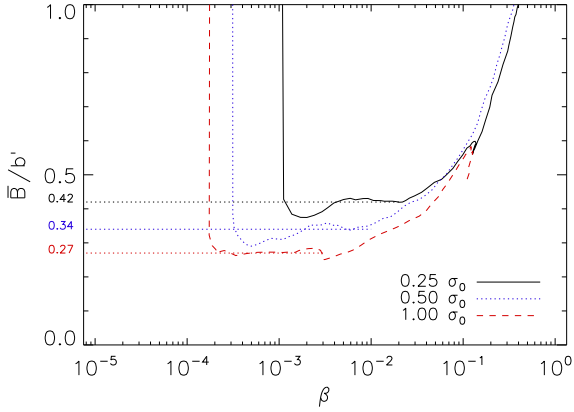}
  \caption{Ratio, $|\mB|/|\mathbf{b}'|$, of the average and turbulent magnetic 
  field strengths versus the relative mean field strength, $\beta$ 
  (formally defined in Eq.~(\ref{eq:beta}) in section \ref{sec:quench} 
  below) for model Q, H, and F. Values are calculated in the vertical 
  range $|\,z,|<1\kpc$.  Within this range, we observe regular vertical 
  profiles for the dynamo tensor coefficients. Flat regions satisfy a 
  scaling relation $|\mB|/|\mathbf{b}'| \sim \sigma^{-0.30 \pm 0.07}$, as 
  also found by \citet{2008A&A...486L..35G}.}
  \label{fig:Bm_Brms}
\end{figure}

Fig.~\ref{fig:Bm_Brms} shows the evolution of the ratio $|\mB| /
|\mathbf{b}'|$ as a function of the relative field strength,
$\beta\equiv \bar{B}/B_{\rm eq}$, where $B_{\rm eq}$ represents the
field strength where the magnetic energy is in equipartition with the
turbulent kinetic energy.\footnote{We note that $\beta$ should not be
  confused with the plasma parameter.} The ratio $|\mB| /
|\mathbf{b}'| $ remains constant during the kinematic phase
(characterised by $\beta\ll 1$), but later in the dynamical phase
(with $\beta \simeq 1$) this ratio increases by the increasing
importance of the background shear term relative to the
turbulence. Typical values of $|\mathbf{b'}|$ at the end of the
kinematic phase are $\sim2$ to $4\muG$, with corresponding values of
$|\mB|/|\mathbf{b}'|$ approximately $0.3$ to $0.5$.  After a short
initial decrease, the ratio remains constant during the kinematic
growth phase of $E_{\rm m}$. The stationary values of this ratio
depend on the SN rate by the relation
$|\mB|/|\mathbf{b}'|\sim\sigma^{-0.30 \pm0.07}$, which matches with
the observations of \citet{2008A&A...482..755C}, and which was also
previously found in similar simulations without a net-vertical flux
\citep{2008A&A...486L..35G}.  During the dynamical phase, `$z$'
profiles of the turbulent field have the same scale heights as that of
$\bar{B}_{y}$, suggesting that magnetic fluctuations are likely
produced via the mechanism of field-line tangling, that is $
\mathrm{b'} \sim \tau_{\rm c}l_{\rm d}^{-1}\bar{B}\,u_{\rm rms}$,
where $\tau_{\rm c}$, and $l_{\rm d}$ are appropriate turbulent
coherence time and diffusion length scales, respectively. We remark
that the field-line tangling term, $\nabla \times
(\mathbf{u}'\times\mB)$ is distinct from the small-scale dynamo term,
represented by $\nabla \times (\mathbf{u}' \times \mathbf{b}')$ in the
induction equation for the fluctuating field, and that it transfers
magnetic energy from ordered ($\mB$) to fluctuating ($\mathbf{b}'$)
fields by ``tangling up'' the large-scale coherent field.  Because, in
the regime of supersonic turbulence, the small-scale dynamo possesses
a higher critical Reynolds number
\citep{2011PhRvL.107k4504F,2014ApJ...797L..19F}, field-line tangling
may provide an attractive alternative to the classic small-scale
dynamo
\citep{1968JETP...26.1031K,2012PhRvE..86f6412S,2012ApJ...754...99S,2013A&A...560A..87S}
in explaining strong fluctuating magnetic fields.

\subsubsection{Alfv\'en velocities}

Vertical profiles of the \emph{total}-field Alfv\'en velocity are
given by the relation
\begin{equation}
  v_{\rm A}(z) = \frac{1}{L_{x} L_{y}}
  \int\int{\frac{|\mathbf{B}|}{\sqrt{\rho}}}\ {\rm d}x\,{\rm d}y\,.
  \label{eq:alf_v_direct}
\end{equation}
The resulting $v_{\rm A}(z)$ saturates to an inverted-bell-shaped
vertical profile for all SN rates. Within the midplane, it scales with
$\sigma$ with amplitudes of $10$, $15$ and $18\kms$ for models Q, H
and F, respectively. Above $\sim 1.0\kpc$, its amplitude ranges up to
$\sim28\pm5 \kms$, irrespective of $\sigma$. Alfv\'en velocities are
typically represented as a ratio of vertical profiles of the rms
magnetic field, $B(z)$, and the square root of the average density,
that is,
\begin{equation}
  v_{\rm A}'(z) = \frac{B(z)}{\sqrt{ \bar{\rho}}}\,.
  \label{eq:alf_v_indirect}
\end{equation}
Table~\ref{tab:b_scales} shows the midplane values of $v_{\rm A}'$. 
These are 10, 14, and $18\kms$ for models Q, H, and F, respectively, and as such roughly scale as $\sigma^{0.4}$ in the midplane.  
The corresponding values in the halo (that is, for $|\,z\,| > 1.0\kpc$) 
are found to be constant with $\sigma$. During the kinematic phase, Alfv\'en velocity profiles $v_{\rm A}(z)$ and $v'_{\rm A}(z)$ differ
by about 25\% in the halo (with $v_{\rm A} < v_{\rm A}'$), but they are 
same within the disc. In contrast, in the dynamical phase 
they match with a good accuracy, except the difference of $\sim-15\%$ 
within the range of $\pm0.8$ to $\pm1.2\kpc$ ($v'_{\rm A} < v_{\rm A}$), 
potentially indicating the loss of a statistical correlation 
between density and Alfv\'en velocity in the dynamical evolution 
phase.

\subsubsection{Mean and fluctuating fluid velocities}
\label{sec:avg_u}
Similar to the magnetic field, velocity fields, $\mathbf{u}$, are 
also split into their mean ($\bar{\mathbf{u}}$) and fluctuating 
part ($\mathbf{u}'=\mathbf{u}-\bar{\mathbf{u}}$). The mean flow 
velocity, $\bar{\mathbf{u}}$, is defined as
\begin{equation}
  \bar{\mathbf{u}}(z) = \frac{1}{L_{x}L_{y}}
                       \int \int{\mathbf{u}}\ {\rm d}x\,{\rm d}y\,.
  \label{eq:mean_vel}
\end{equation}
Since there is no radial (or azimuthal) variation of SN distribution, 
only the ${\bar{u}}_z(z)$ profiles are non-zero, and we find them to 
be roughly linear in $z$ (see Fig.~\ref{fig:quench}, {\it lower-left}, black solid lines). For their overall amplitude, we observe a power 
law scaling with respect to SN rate as ${\bar{u}}_z(z)\sim
\sigma^{0.4}$ (also cf. Table~\ref{tab:alphas}, below). As a consequence of 
the vertical density stratification, $\mathbf{u}'(z)$ has an inverted-bell-shape profile, (similar to the Alfv\'en velocity) within the 
inner disc ($|\,z\,| < 0.8\kpc$), and a constant or decreasing linear 
profile within the outer halo ($|\,z\,| > 0.8$). The resulting 
M-shaped  profile illustrates that, despite the energy input is 
peaking in the midplane, it is easier to maintain a velocity 
dispersion at intermediate densities than within the dense Galactic 
midplane. The maxima of $\mathbf{u}'$  (which are situated at $z=
\pm0.8\kpc$) are 20, 30 and $40\kms$ for models Q, H and F, respectively and scale roughly with the SN rate as $\mathbf{u}' 
\sim\sigma^{0.4}$. The midplane 
values of 6, 11 and 25 $\kms$ (for Q, H and F), however, scale as $\sim\sigma^{1.0\pm0.1}$. 
In conclusion, we also remark that the width of the 
inner $\mathbf{u}'(z)$ profile becomes broader ($-1.0\kpc$ to $+1.0
\kpc$) in the dynamical growth phase of $E_{\rm mag}$.

\subsection{Growth and saturation of the dynamo} 
\label{sec:dynamo}

In order to understand the kinematic and dynamic phases of magnetic
energy amplification for different SN rates seen in
Fig.~\ref{fig:tmag}, we employ a mean-field approach based on the
$\alpha\Omega$~dynamo framework. A similar model has been used
previously to explain the initial amplification of the magnetic energy
in the kinematic evolution phase \citep{2010PhDT.......244G} of the
dynamo. For the present analysis, we use the standard mean-field
formulation \citep[see, for instance,][]{2007mhet.book...55R}. The
approach is based on obtaining a closure for the correlation between
the turbulent velocity ${\mathbf{u}}'$ and turbulent magnetic field
$\mathbf{b}'$ -- the so-called `turbulent electromotive force',
$\mathcal{E}$, which is itself a mean quantity:
\begin{equation}
  \mathcal{E} = \overline {\mathbf{u}' \times \mathbf{b}'}\,.
  \label{eq:mean_emf}
\end{equation}
We adopt a local mean-field formulation in Cartesian geometry,
\citep{2005AN....326..787B}, and define the average of the
fluid variables ($\mathbf{B}$, $\mathbf{u}$) over the ${xy}$
plane (cf. section~\ref{sec:avg_b} and section~\ref{sec:avg_u}). 
Motivated by the second-order correlation approximation (SOCA), the 
vertical profile, $\mathcal{E}(z,t)$ is expanded into a
linear function of the average magnetic field profile,
$\bar{\mathbf{B}}(z,t)$ as
\begin{equation}
  \mathcal{E}\,(z,t) = \alpha\,\mB\,(z,t)
                     - \eta\,\nabla\times\mB\,(z,t)\,,
\label{eq:mean_emf_dyn}
\end{equation}
where $\alpha$ and $\eta$ are now tensorial quantities\footnote{Note
  that $\eta$ here is different from the microscopic \emph{scalar}
  diffusivity, $\eta_{\rm m}$ introduced in
  Equation~(\ref{eq:mhd}). Because $\eta_{\rm m} \ll |\eta|\,$, the
  microscopic diffusion can safely be neglected in the mean-field model.}  referred
to as the `dynamo coefficients'. Applying the SOCA approximation to
isotropic, homogeneous turbulence, one can associate the diagonal
elements of the $\alpha$~tensor with kinetic helicity, $\alpha_{xx} =
\alpha_{yy} = \alpha_{\rm iso}\sim -\tau_{\rm c}\,\overline{
  \mathbf{u}' \cdot \left(\nabla\times\mathbf{u}'\right)}/3$ (note the
minus sign).  Another important effect known from the realm of this
theory, is the turbulent (or \emph{diamagnetic}) pumping, $\gamma
\equiv 0.5\,(\alpha_{xy} - \alpha_{yx})$, which describes the
non-advective redistribution of magnetic flux by gradients in
turbulent intensity, that is, $\gamma\sim -\tau_{\rm
  c}/6\,\nabla\mathbf{u}'^2$, again from SOCA. The diagonal elements
of the $\eta$~tensor are commonly referred to as ``eddy'' diffusivity,
and the SOCA estimate is $\eta \sim \tau_{\rm
  c}/3\,\mathbf{u}'^2$. Note that the quantities $\tau_{\rm c}$
appearing in the preceding expressions are not required to be
identical but may differ by factors of order unity among each other.

\subsubsection{Quenching of mean-field coefficients}
\label{sec:quench}

To compute approximate values of $\alpha(z,t)$ and $\eta(z,t)$ 
tensors from our DNS, we use the test-field approach as described in 
\citet{2008A&A...482..739B}. To obtain enough independent pieces of 
information to invert the tensor equation (\ref{eq:mean_emf_dyn}), 
additional \emph{passive} test fields are evolved parallel to the 
main simulation run. $\mathcal{E}(z,t)$ is then computed for the 
associated test-field fluctuations, $\mathbf{b}'$, via Eq.~
(\ref{eq:mean_emf}), and Eq.~(\ref{eq:mean_emf_dyn}) is inverted to 
obtain the tensors $\alpha(z,t)$ and $\eta(z,t)$ as a function of 
time. A detailed description of this method is beyond the scope 
of this paper, and we refer the interested reader to appendix~B in 
\citet{2010PhDT.......244G}.

\begin{figure*}
  \centering
  \includegraphics[width=0.8\columnwidth]{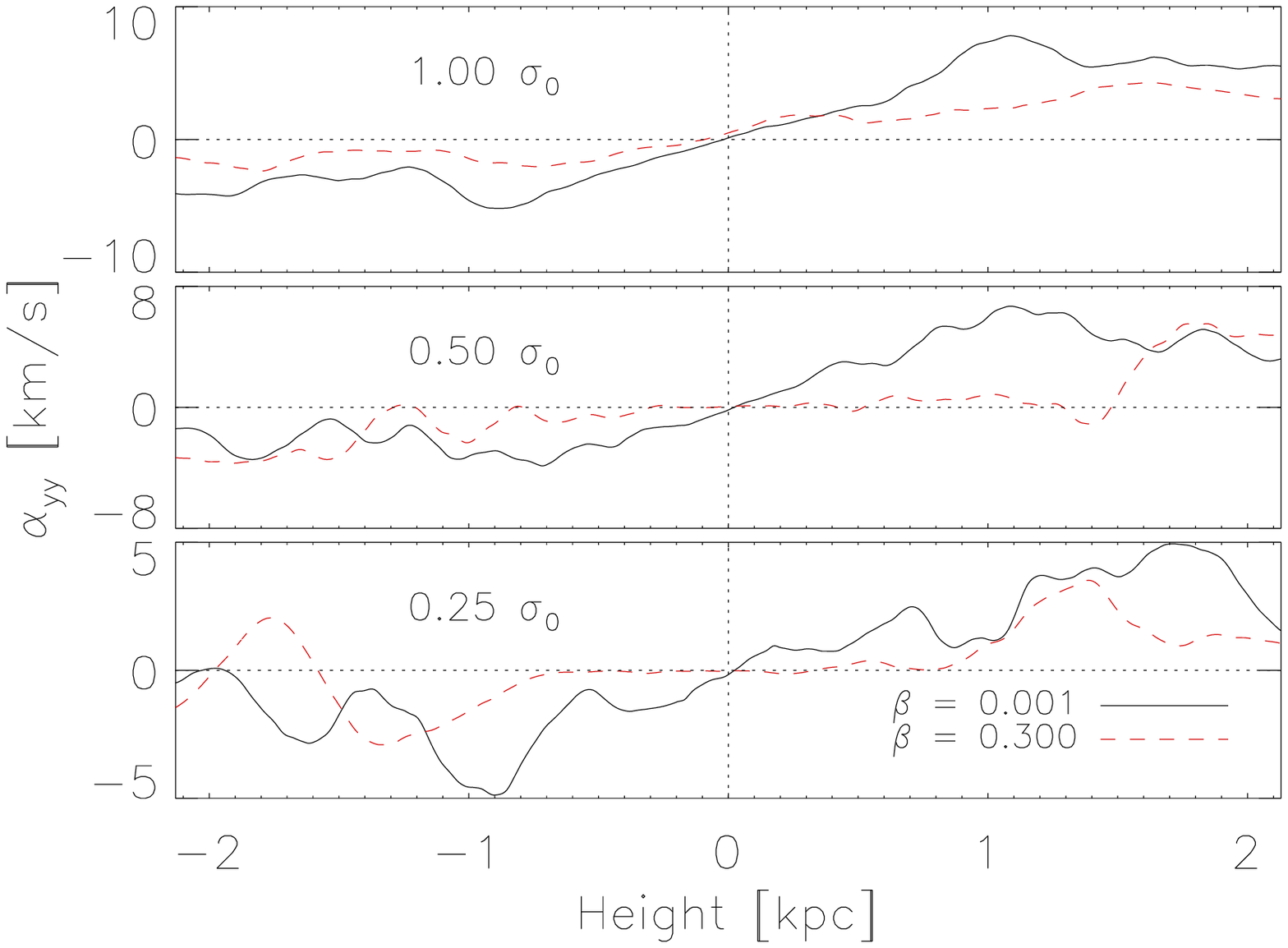}\hspace{20pt}
  \includegraphics[width=0.8\columnwidth]{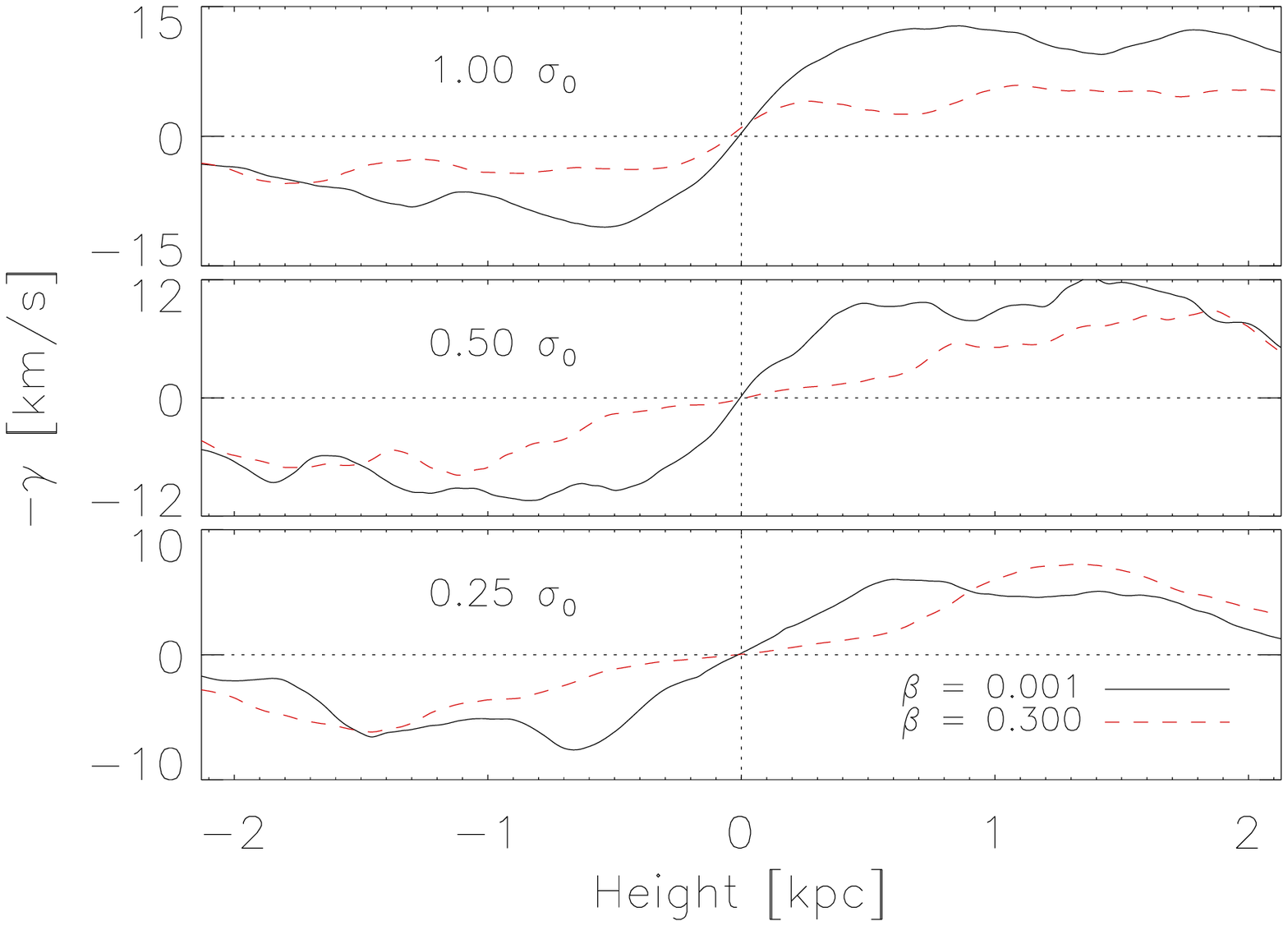}\\[10pt]
  \includegraphics[width=0.8\columnwidth]{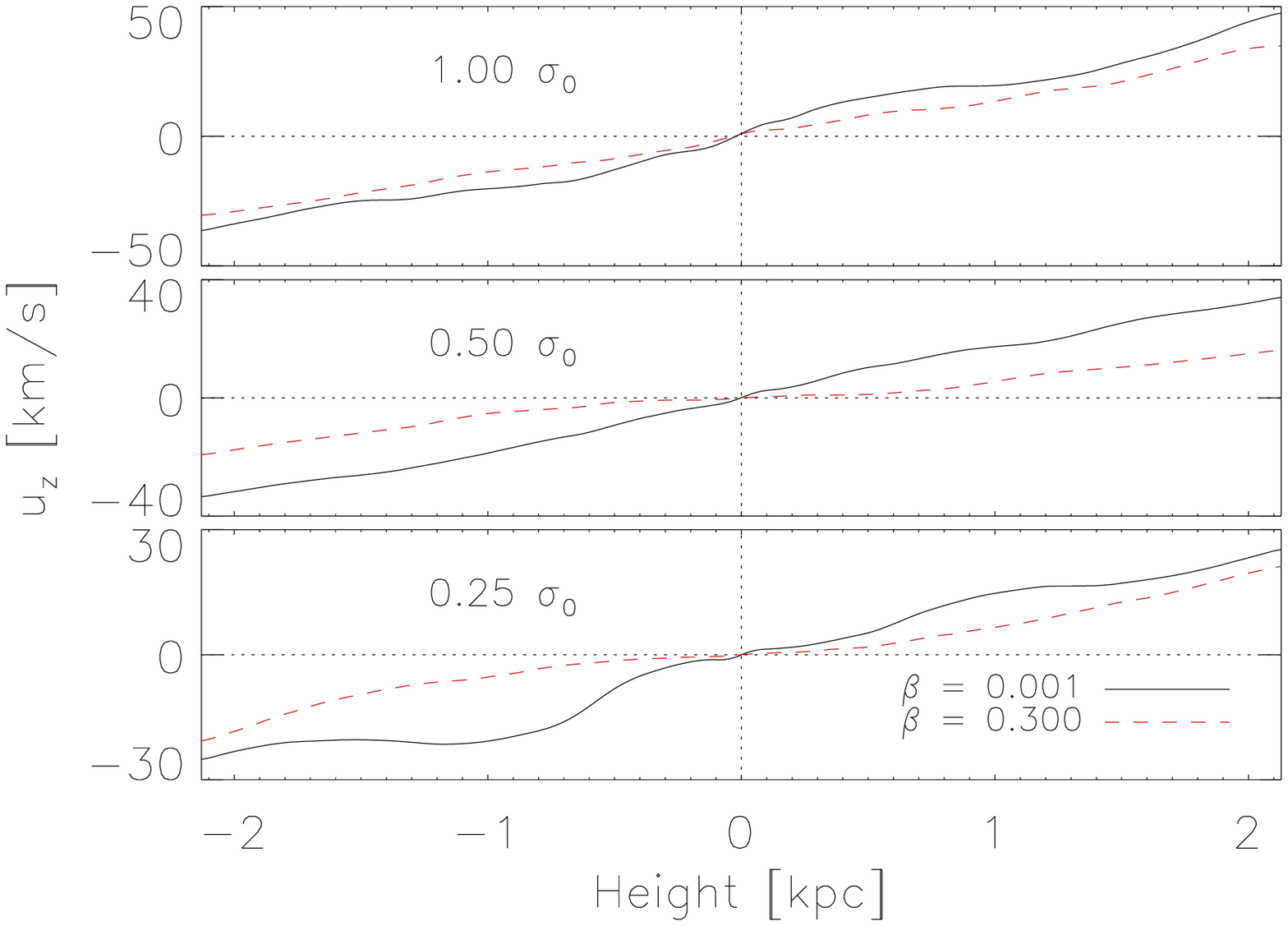}\hspace{20pt}
  \includegraphics[width=0.8\columnwidth]{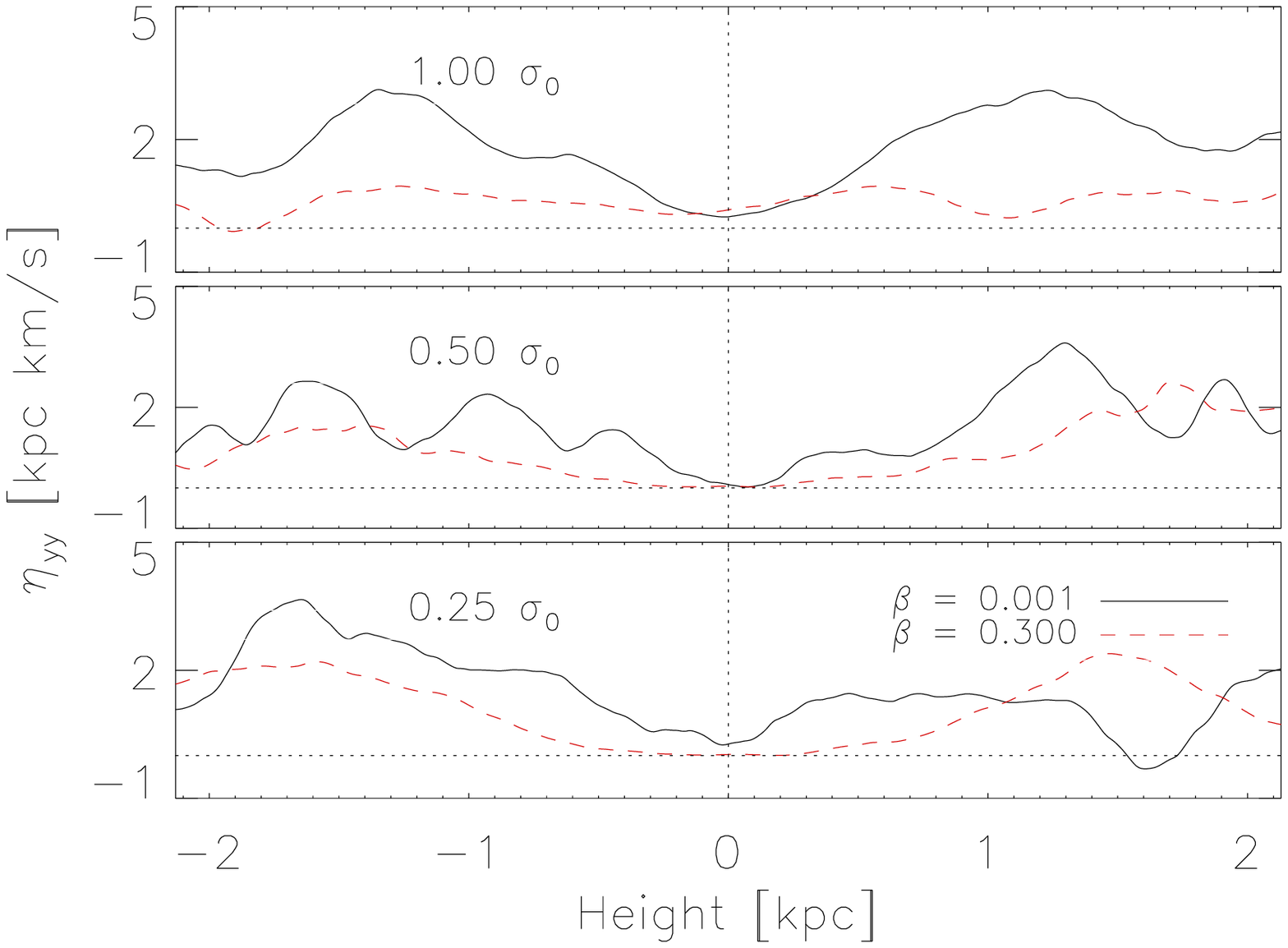}
  \caption{Vertical profiles of the various dynamo coefficients of 
  interest, averaged over $0.2\Gyr$ in the kinematic phase (solid 
  black lines) and in the dynamic phase (dotted red lines). The three 
  plots of each sub-panel correspond to model F, H, and Q, with 
  diminishing supernova rate, $\sigma=1$, 0.5, and 0.25, respectively. 
  The quantities shown are (top-left to bottom-right): the dynamo 
  effect, $\alpha_{yy}(z)$ (relevant for the $\alpha\Omega$ mechanism), diamagnetic pumping, $\gamma(z)$ (note the negative sign), the mean 
  vertical flow velocity, ${\bar{u}}_z(z)$, and the eddy diffusivity, 
  $\eta_{yy}(z)$. The values of the corresponding normalised field 
  strength, $\beta$, defined by Eq.~(\ref{eq:beta}), are $10^{-3}$ 
  (kinematic regime) and $0.3$ (quenched regime), respectively.
    \label{fig:quench}}
\end{figure*}

The black solid lines in Fig.~\ref{fig:quench} show the $z$ profiles
of all dynamo coefficients for three different values of $\sigma$,
corresponding to model, F,H, and Q, respectively. We also list, in
Tab.~\ref{tab:alphas}, the values of $\alpha$ and $\eta$ in the halo
for the different values of $\sigma$ in the \emph{kinematic} phase.
The amplitude of the dynamo coefficients is found to scale with the
supernova rate as $\sim\sigma^{0.4}$ (see Table~\ref{tab:alphas}).  In
reality, the increase in the SN rate is associated with the increase
in density \citep[e.g.][]{1998ApJ...498..541K}. For the present
analysis, however, we do not change these two parameters
simultaneously. This is done in order to obtain the explicit
dependencies on these variables, which can later be used in
parametrisations.

\begin{table}
  \caption{Dynamo coefficients (at $z=1\kpc$), and vertical wind
    velocity $\bar{u}_z$ (at $z=2\kpc$), both in the kinematic
    phase. Scaling laws for the respective amplitudes of
    $\alpha$, $\eta$, $\gamma$, and $\bar{u}_z$ are $\sim
    {\sigma}^{0.4}$.}\centering
  \begin{tabular}{lcccc}
    \hline 
    & $\sigma$ & $\alpha_{xx}$ & $\alpha_{yy}$  & $\gamma$ \\
    & $[\sigma_0]$ & $[\kms]$ & $[\kms]$ & $[\kms]$ 
    \\[2pt]\hline\\[-6pt]
    Q & 0.25 & $3.5\pm0.8$ [4.0] & $3.8\pm0.9$ [4.5] & $9.7\pm2$ [7.6]\\
    H & 0.50 & $4.6\pm1.7$ [5.3] & $4.9\pm2.0$ [6.6] & $ 12\pm2$ [9.5]\\
    F & 1.00 & $6.1\pm2.0$ [7.0] & $6.6\pm3.0$ [7.8] & $ 15\pm4$ [13.5]\\[4pt]
    \hline 
    & $\sigma$ & $\eta_{xx}$ & $\eta_{yy}$ & $\mathrm{u_z}$ \\
    & $[\sigma_0]$ & $[\kpc\kms]$ & $[\kpc\kms]$ & $[\kms]$ 
    \\[2pt]\hline\\[-6pt]
    Q & 0.25 & $1.8\pm0.5$ [1.5] & $2.6\pm0.6$ [2.1] & $ 28\pm2$ [30] \\ 
    H & 0.50 & $2.5\pm0.8$ [1.9] & $3.3\pm1.1$ [2.7] & $ 35\pm4$ [37] \\ 
    F & 1.00 & $3.2\pm1.5$ [2.5] & $4.4\pm1.6$ [3.1] & $ 45\pm4$ [44] \\ \hline
  \end{tabular}\\[8pt]
    \parbox[t]{\columnwidth}{\footnotesize \emph{Notes}: Numbers in 
    the squared bracket represent the values of corresponding 
    coefficient, obtained by fitting the data with the Legendre 
    polynomials, which we further use in 1D simulations  (cf. 
    section~\ref{sec:mf_dynamo}).}
  \label{tab:alphas}
\end{table}

The dynamo coefficients are found to be quasi-stationary in time only
while the flow structure is uninfluenced by the magnetic field, that
is, in the kinematic phase, characterised by $v_{\rm A} \ll u'$. In
the dynamical phase, the expansion of $\mathcal{E}$ from
Eq.~(\ref{eq:mean_emf_dyn}) is no longer a strict linear function of
$\bar\mathbf{B}$, but is non-linearly ``quenched'' \citep[see,
  e.g.,][]{2007mhet.book...55R}. As a consequence, in the most simple
case, the amplification of $\bar\mathbf{B}$ is expected to slow down
if the $\alpha$~effect diminishes more strongly than the turbulent
diffusivity, $\eta$.

The quenching of the vertical profiles is qualitatively 
represented in Fig.~\ref{fig:quench}, where black-solid lines show 
the initial unquenched profiles, which during the dynamical evolution 
stage become less steep (as shown by red dashed lines). To describe 
this effect in a quantitative manner, we here represent the dynamical 
importance of mean fields by the ratio $\beta$, which describes the 
relative magnitude of the Alfv\'en and turbulent rms velocities, that 
is,
\begin{equation}
  \beta =
  \frac{{\left(\bar{B}_x^2+\bar{B}_y^2\right)}^{1/2}}{\sqrt{\rho\,u'^2}}
  \,,
\label{eq:beta}
\end{equation}
equivalent to the square-root of the ratio of magnetic and kinetic
pressure, and where we ignore the contribution from the (fixed) weak
vertical mean field. Since the vertical profiles of turbulent kinetic
energy and turbulent magnetic fields are Gaussian, the vertical
profile of $\beta$ has the same functional form, i.e.  $\beta$ in the
disc is greater than its halo value by $\sim20\%$. We note that, when
increasing the supernova rate, the vertical profiles of
$\beta\left(z\right) $ become flat. For model F, the difference
between the disc and halo values of $\beta$ is $\sim5\%$, while for
model Q, the difference becomes as high as $\sim30$ to $40\%$.

To derive the dependence of dynamo effects on the presence of mean fields, 
we fit the coefficients by a standard algebraic function of $\beta(z)$ 
as defined by, eq.~(1) in \citet{2013MNRAS.429..967G}. Recently, 
\citet{2014MNRAS.443.1867C} used a similar approach to demonstrate the 
saturation of $E_{\rm m}$ in 1D dynamo models. Using the test-field 
method, we obtain the following quenching relations 
\citep{2013MNRAS.429..967G}\smallskip
\begin{equation}
  \alpha = \frac{\alpha_{\rm k}}{1+27\,\beta^{2}}\,,\quad
  \eta   = \frac{\eta_{\rm k}}{1+6\,\beta}\,,\quad
  \gamma = \frac{\gamma_{\rm k}}{1+10\,\beta^{2}}\,,
  \label{eq:dyn_coeff}
\end{equation}
where $\alpha_{\rm k}$, $\eta_{\rm k}$ and $\gamma_{\rm k}$ are 
the \emph{unquenched} amplitudes of the dynamo coefficients during 
the initial kinematic phase, that depend upon the SN rate as 
$\alpha_{\rm k} = \alpha_0\ (\sigma/\sigma_0)^{0.4}$, $\eta_{\rm k}
=\eta_0\ (\sigma/\sigma_0)^{0.4}$, and $\gamma_{\rm k}=\gamma_0\ 
(\sigma/\sigma_0)^{0.4}$ (see Table~\ref{tab:alphas}).

\begin{figure}
\centering
\includegraphics[width=\columnwidth]{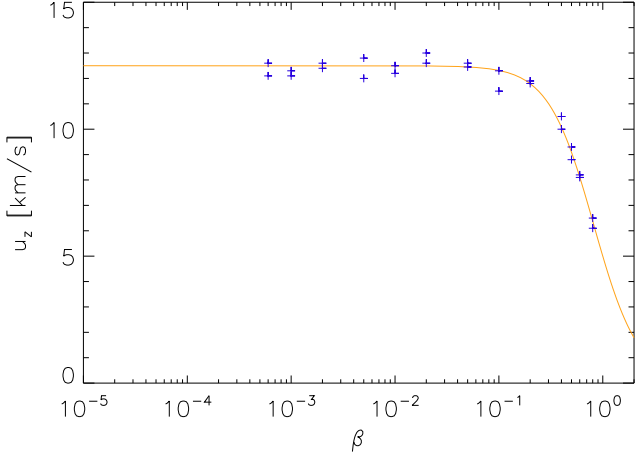}
\caption{Quenching of the vertical wind with respect to $\beta$, 
calculated for model Q at a reference height of $z=\pm0.8\kpc$, 
representing the vertical disc region that we find to be affected by 
the quenching. The resulting relation, $\bar{u}_z = \bar{u}_{z\,0}/
{(1+1.5\,\beta^{2})}$, means that quenching of the wind sets in at a 
comparatively higher value of $\beta$ than in relation 
(\ref{eq:dyn_coeff}) for $\gamma(z)$.  Two '+' signs corresponding to 
the same $\beta$ represent the value of $\bar{u}_z$ above and below 
the midplane}
\label{fig:uz_quench}
\end{figure}

The horizontal components of the mean velocity in our co-rotating 
frame are negligibly small, so we have to include only the mean 
vertical velocity $\bar{u}_z$ into the dynamo equations. The 
amplitude of the vertical profile of the mean velocity was found 
to scale with the supernova rate as $\sigma^{0.4}$, (see Fig.~\ref{fig:quench} and Tab.~\ref{tab:alphas}). In the dynamical phase, 
the mean velocity profiles furthermore undergo non-linear quenching.  
We find the best fit for it using our direct simulations data, which 
has the same algebraic form as for $\gamma(\beta)$. The quenching for 
model Q and H appears mainly around the midplane, where $\beta$ is 
maximal.  Figure~\ref{fig:uz_quench} shows the dependence of 
${\bar{u}}_z$ on $\beta$ at $z = \pm0.8\kpc$. This dependence 
is also reflected in the linear slope of the $\bar{u}_z(z)$ profile 
within approximately $-1\kpc<z<+1\kpc$, which is to say that we 
would obtain a similar $\beta$ dependence if we were to measure this 
relation at a different reference point within this interval. The 
obtained quenching of ${\bar{u}}_z$ is
\begin{equation}
  \bar{u}_z =  \frac{\bar{u}_{\rm k}}{1+1.5\,\beta^2}\,,
  \label{eq:wind_quen}
\end{equation}
where the initial unquenched amplitudes $\bar{u}_k$ also scale with 
the SN rate as $\bar{u}_k = \bar{u}_0\ (\sigma/\sigma_0)^{0.4}$ (see 
Tab.~\ref{tab:alphas}).

Above $\sim1.2\kpc$, fluctuations in $\bar{u}_z(z)$ are comparatively
larger, and the estimated error in $\bar{u}_z(\beta)$ from
Eq.~(\ref{eq:wind_quen}) is more than $\pm40\%$.
Equation~(\ref{eq:wind_quen}) also enters the mean-field description
of the one-dimensional $\alpha\Omega$~dynamo model that we will
describe in the following section. The quenching of the wind profiles
can be qualitatively seen in the lower-left panel of
Fig.~\ref{fig:quench}, and is comparatively weaker than the $\gamma$
quenching -- as can be seen from Eqs.~(\ref{eq:dyn_coeff}) and
(\ref{eq:wind_quen}).

\subsubsection{The mean-field dynamo model}
\label{sec:mf_dynamo}

With the average defined over the plane coordinates $x$ and $y$, we 
obtain the set of 1D dynamo equations
\begin{eqnarray}
  \frac{\partial{\bar{B}_{x}}}{\partial t} &=&
  \frac{\partial}{\partial
    z}\,{\left(-\left({\bar{u}}_z+\gamma\right)\,\bar{B}_{x}
    -\alpha_{{yy}}\,\bar{B}_{y} +
    \eta_{yy}\,\frac{\partial{\bar{B}_{x}}}{\partial z}\right)}\,,
  \nonumber\\[4pt]
  \frac{\partial{\bar{B}_{y}}}{\partial t} &=& 
  \frac{\partial}{\partial z}\,{\left(-\left({\bar{u}}_z+\gamma\right)
    \,\bar{B}_{y}
    +\alpha_{{xx}}\,\bar{B}_{x} +
    \eta_{xx}\,\frac{\partial{\bar{B}_{y}}}{\partial z}\right)}
    + q\,\Omega \bar{B}_x\,,
  \nonumber\\[4pt]
  \frac{\partial{\bar{B}_{z}}}{\partial t} &=& 0\,,
    \label{eq:dyn_eqn}
\end{eqnarray}
where we choose to drop the last equation.  We furthermore neglect
unimportant small contributions from the off-diagonal elements of the
$\eta$~tensor, as well as any symmetric contributions to the
off-diagonal elements of the $\alpha$~ tensor. We obtain
approximations of the unquenched functions $\alpha_0(z)$, $\eta_0(z)$
and $\gamma_0(z)$ from DNS, using the test-field method, and average
over the first $200$ to $400\Myr$ to reduce stochastic
contributions. Furthermore, for the purpose of filtering
high-wavenumber fluctuations, we expand the $\alpha_0$, $\gamma_0$
($\eta_0$) profiles into a series of odd (even) Legendre polynomials,
$P_n(z)$, up to order $n=15\, (14)$. This approach is equivalent of
applying a low-pass filter, and at the same time enforces the correct
(odd/even) symmetry with respect to $z$. It turns out that, because of
its simple shape, the wind, that is $\bar{u}_z(z)$, is sufficiently
fitted by the linear first order Legendre polynomial. We close the
system by the derived quenching formulas Eqs.~(\ref{eq:dyn_coeff}) and
(\ref{eq:wind_quen}). Here we use the plane averaged turbulent kinetic
energy to calculate $\beta$ defined by Eq.~(\ref{eq:beta}). Because of
the negligible time dependence of the turbulent kinetic energy
throughout the evolution for all SN rates (see
section~\ref{sec:evol}), we use a function for $B_{\rm eq}(z)$
averaged over the first $300$ to $400\Myr$.

We discretise the resulting set of partial differential equations via 
a finite difference approach on a staggered grid. We apply similar 
boundary conditions for $\bar{B}_{x}$, $\bar{B}_{y}$ as in the direct 
simulations. To facilitate a direct comparison, we furthermore choose 
the initial profiles $\bar{B}_{x}$, $\bar{B}_{y}$ from the DNS data, 
averaged over first $50\Myr$ to $100\Myr$. This ensures that the 
mean-field simulations are seeded with the same mix of modes that develop 
self-consistently in the early evolution of the DNS.

The diamagnetic pumping term, $\gamma$, appears in the evolution 
equation for the mean fields in the form of $\bar{u}_z + \gamma$. 
In the kinematic phase these two terms balance each other within 
the disc (see Fig.~\ref{fig:quench}, black solid lines in the 
upper right and lower left panels, respectively). In the absence 
of dominant transport processes, the $\alpha\Omega$~dynamo operates 
in an efficient regime, leading to an exponential amplification 
of mean field via Eq.~(\ref{eq:dyn_eqn}) for all SN rates.

\begin{figure}
  \centering
  \includegraphics[width=\columnwidth]{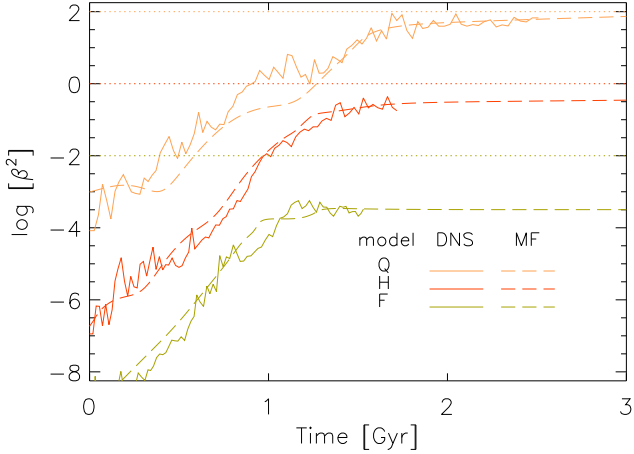}
  \caption{Evolution of the magnetic energy for the mean fields
    (normalised with the turbulent kinetic energy), i.e. $\beta^2$,
    from direct numerical simulations (DNS) and 1D dynamo simulations
    (MF). Different colours indicate the different supernova
    rates. The curves for model Q and F are shifted upwards and
    downwards, respectively, likewise the dotted horizontal lines at
    $log(E_m/E_k)=2$, $0$ and $-2$ mark equipartition for models Q, H
    and F respectively. The growth and the saturation of the magnetic
    energy seen in the DNS is effectively reproduced via the 1D dynamo
    model using $\alpha$, $\eta$ and $\bar {u}_z$ with the quenching
    via Eqs.~(\ref{eq:dyn_coeff}) and (\ref{eq:wind_quen}).}
  \label{fig:dns_dyn_comp}
\end{figure}

In the dynamical phase, due to the flattened profiles of $\alpha$ (and
$\eta$) the amplification process slows down in the Q and H models,
while it stops in model F. As illustrated in
Fig.~\ref{fig:dns_dyn_comp}, where we compare the evolution of the
magnetic energy in the direct simulations with the 1D dynamo model,
the simple dynamo model does a reasonable job in reproducing the field
evolution seen in DNS. Unlike in models Q and H, in model F, a
non-vanishing contribution of the (${\bar{u}}_z+\gamma$) term is found
to become important for comparatively smaller values of $\beta$ and
the amplification of $\mB$ is inhibited as a result. 

The sustained growth of the magnetic field appears to be a consequence 
of the combined quenching of $\alpha$ and $\eta$ -- leading to an 
indefinitely growing field. This is also witnessed by the 
dynamo number $D$ (defined in Eq.~(\ref{eq:dyn_def}) below), 
that becomes a function of $\beta$ according to Eq.~(\ref{eq:dyn_number}). Only the growth time scale for the dynamo 
is reduced by the quenching of the diffusivity. This behaviour is 
clearly reflected in the 1D models for the cases Q and H, that 
show much reduced growth at late times. In contrast, in model F, at 
higher supernova rate, both the pumping, $\gamma$, and the wind, 
$\bar{u}_z$, appear to be less affected by the buildup of strong 
magnetic fields (see Fig.~\ref{fig:quench}). In this situation it 
appears probable that the $\alpha\Omega$~dynamo is instead 
saturated by advective losses via the residual transport velocity 
($\bar{u}_z+\gamma$). We have tested this scenario by switching 
off the transport term in all three models, which resulted in an 
endlessly growing field also for model~F. We analytically justify 
this result in the following section.  The effect of pumping and 
advective losses relative to the propagation direction of the 
dynamo wave has previously been studied in the kinematic phase 
only \citep[see fig.~3 in][]{2011EAS....44...73G}.

The ability to reproduce the evolution of $\bar{B}_x(z)$ and
$\bar{B}_y(z)$ profiles (as shown in the lower right panel in
Fig.~\ref{fig:mean_fields}) using a simple one-dimensional
$\alpha\Omega$~dynamo is rather remarkable. The space-time plots in
this figure compares the evolution of the mean azimuthal field from 1D
dynamo simulations (upper panel) and from DNS (lower panel) for model
Q. The vertical scale heights and midplane field strengths of
$\bar{B}_x(z)$ are also comparable for both of these simulations --
see panels (a) and (b) in Fig.~\ref{fig:mean_fields}.

\subsubsection{Assessment in terms of dynamo numbers}

The time where the growth rate of $E_{\rm m}$ changes is clearly seen 
in Fig.~\ref{fig:tmag}. For instance, $E_{\rm m}$ saturates after 
$1.2$, $1$ and $0.9\Gyr$ for models Q, H and F, respectively. 
We were able to analytically derive 
the value of the total magnetic energy at the time where it changes 
its slope, and this is done as follows: The dynamo number for the 
$\alpha\Omega$~dynamo is defined as \citep[e.g.][]{1988Natur.336..341R},
\begin{equation}
  D \equiv C_\alpha\times C_\Omega 
  = \frac{\alpha H}{\eta}\times \frac{\Omega H^2}{\eta}
  = \frac{\alpha\ H^{3} \Omega}{\eta^{2}}\,,
  \label{eq:dyn_def}
\end{equation}
where $C_\alpha$, and $C_\Omega$ denote the contributions from the 
$\alpha$~effect and the shear, respectively, and where $H$ is a typical 
vertical length scale. Substituting Eq.~(\ref{eq:dyn_coeff}) into Eq.~(\ref{eq:dyn_def}), we yield an expression for this number in terms of 
the supernova rate, $\sigma$, and the relative magnetic field strength, 
$\beta$, that is,
\begin{equation}
  D = D_{0}\,\left(\frac{\sigma}{\sigma_0}\right)^{-0.4}\,
                   \frac{\left(1+6\,\beta\right)^{2}}{1+27\,\beta^{2}}\,,
\label{eq:dyn_number}
\end{equation}
where $D_0$ is the \emph{unquenched} dynamo number obtained
with values for $\alpha$, and $\eta$ in the kinematic phase.

Assuming $\bar{u}_z(z) \simeq \gamma(z)$, and using the known
approximate solution for the $\alpha\Omega$~dynamo for stationary
profiles of the dynamo coefficients \citep[similar
  to][]{1998MNRAS.299L..21S}, we obtain 
\begin{equation}
  \bar{B}(t) = \bar{B}_{0}\ \mathrm{exp}\left( {\frac{t}{t_{\rm d}}
    \sqrt{\frac{D}{D_{\rm c}}-1}\,} \right)\,,
\label{eq:b_evol}
\end{equation}
as an approximate solution expressed in terms of a diffusion time 
$t_{\rm d} \equiv H^2/\eta = \sigma^{-0.4}H^{2}\ (1+6\,\beta)/ 
\eta_{0}$, derived from Eq.~(\ref{eq:dyn_coeff}), and 
introducing the critical dynamo number, $D_{\rm c}$, delineating 
marginal growth of the dynamo.  We choose $L = 1\kpc$, and $\eta' 
= 1\kpc\kms$ to normalise $H$ and $\eta_0$ as $H = H'L$ and $\eta_0 
= \widetilde\eta\eta'$.  Using the appropriate relation 
(\ref{eq:dyn_number}) for the dynamo number, we further yield 
$\,\bar{B}(t) = \bar{B}_{0}\ \mathrm{exp} \left({f(\beta,\sigma)\,t}
\right)\,$ with the function $f(\beta,\sigma)$ given by
\begin{equation}
  f(\beta,\sigma) = \frac{\eta_{0}}{H^2}\ \sqrt{\frac{D_{0}}{D_{\rm c}}
    \frac{\sigma^{0.4}}{\left(1+27\beta^2\right)} 
    - \frac{\sigma^{0.8}}{\left(1+6\beta\right)^{2}}}\,.
  \label{eq:f_beta}
\end{equation}
\begin{figure}
  \centering
  \includegraphics[width=\columnwidth]{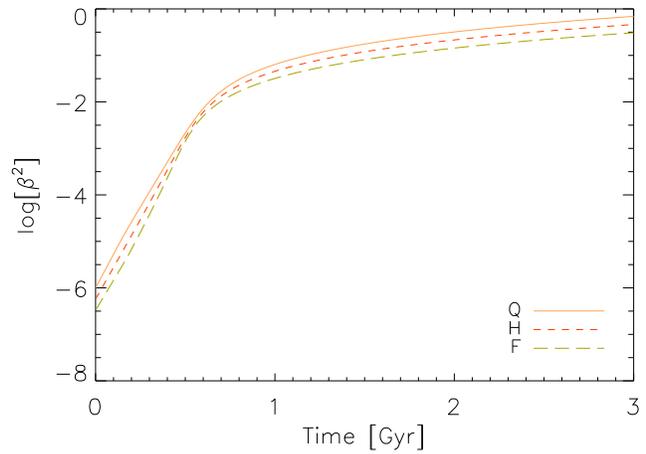}
  \caption{Evolution of the square of relative field strengths, 
  $\beta^{2}$, for different SN rates calculated using Eqs.~(\ref{eq:b_evol}) and (\ref{eq:f_beta}). The e-folding time 
  of the initial kinematic phase (up to $\sim 1\Gyr$) is 
  approximately $100\Myr$ (consistent with the DNS and the 1D-MF 
  simulations).}
  \label{fig:beta_evol}
\end{figure}
This relation provides an evolution equation for the mean magnetic
field. We plot the resulting time evolution of square of the relative
magnetic field strength, $\beta^{2}$, in Fig.~\ref{fig:beta_evol}.
Here, length scales are normalised with $H = 1\kpc$, the ratio $D_{0}
/D_{c} = 3.5$ is obtained empirically from a set of 1D simulations,
and $\eta_{0} = 4.4\kpc\kms$ is taken from Table~\ref{tab:alphas}.
The values of $\bar{B}_0$ and the equipartition magnetic field are
chosen according to the initial values of $\beta$ taken from the DNS,
that is $\beta = 6\times10^{-4}\,\sigma^{-0.4}$. In particular, the
factor $\sigma^{-0.4}$ derives from the square root of turbulent
kinetic energy, $\rho^{0.5}\ u'$, appearing in the denominator of
Eq.~(\ref{eq:beta}). Unlike in the DNS, all three curves show a very
similar behaviour.  This is because the dynamical phase is not well
reproduced in Eq.~(\ref{eq:f_beta}), since the quenching of the wind is
not considered in our analytic description given by
Eq.~(\ref{eq:b_evol}). Model Q (solid line), however, approximately
mimics the entire evolution curve of $\beta$ seen in the DNS,
suggesting that the effect of the outflow is less pronounced at low
supernova rates.

To convert Eq.~(\ref{eq:b_evol}) into an evolution equation for 
the total magnetic energy, we use a relation between the average and 
turbulent fields as a function of $\sigma$ (also see 
Fig.~\ref{fig:Bm_Brms}):
\begin{equation}
  |\mB| = 0.27 \sigma^{-0.30 \pm 0.07}\,|\mathbf{b}'|\,,
  \label{eq:bm_brms_t}
\end{equation}
and express the total magnetic energy as a sum of the magnetic energy 
from the mean and turbulent fields. We finally obtain an evolution 
equation for the total magnetic energy,
\begin{equation}
  E_{\rm m} = E_{0} \left(1+13.7\sigma^{0.6}\right)\,
                    \mathrm{exp}\left(\;{2f(\beta, \sigma)\,t}\;\right)\,.
  \label{eq:em_evol}
\end{equation}
This evolution equation is valid for the initial kinematic phase of 
$E_{\rm m}$, that is, where Eq.~(\ref{eq:bm_brms_t}) is valid.  
Substituting the values for $\sigma$ and $f(\beta, \sigma)$, for 
fixed $\beta$, in Eq.~(\ref{eq:em_evol}), we obtain a scaling for the 
total magnetic energy $E_{\rm m}$ at the end of the kinematic phase. 
This estimate successfully predicts the values obtained in DNS.

\begin{figure}
  \centering
  \includegraphics[width=\columnwidth]{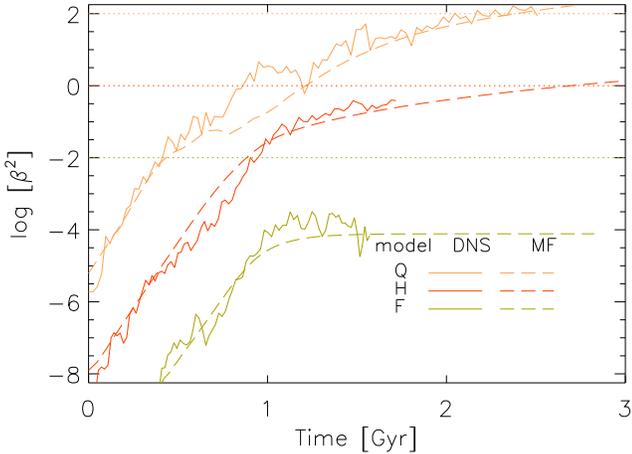}
  \caption{Same as Fig.~\ref{fig:dns_dyn_comp}, except the energies
    from DNS are computed within the inner disc ($|z| < 0.5\kpc$), and
    the 1D mean-field simulations are preformed with the profiles of
    the dynamo coefficients restricted to the thin disc. The e-folding
    time for all models in the initial kinematic phase is $\sim70\Myr$.}
  \label{fig:dns_dyn_comp_disc}
\end{figure}

As is clear from Figs.~\ref{fig:dns_dyn_comp} and \ref{fig:beta_evol},
the $\beta$ values in any of our models do not exceed unity. This is
partly because we have taken averages over the entire volume of the
simulation domain. If we restrict ourselves to the thin disc
($|z|<0.5\kpc$), it is possible to obtain $\beta$ values exceeding
unity, at least for smaller SN rates. This hypothesis is tested with
1D dynamo simulations using vertical profiles of $\alpha$, $\eta$,
$\gamma$ and $\bar{u}_z$ that have been obtained within the range
$-0.5\kpc < z < +0.5\kpc$. Using these, we obtain $\beta^2\sim 1.2$,
$0.35$ and $0.03$ for model Q, H and F as shown in
Fig.~\ref{fig:dns_dyn_comp_disc}, where we plot magnetic energies
computed from the thin disc. We also obtain faster growth times of
$\sim70\Myr$ for $\beta^{2}$ during the initial kinematic phase. The
analytical solution given by Eq.~(\ref{eq:b_evol}), now using $H=0.25
\kpc$ to account for the more localised dynamo mode, also yields {\bf
  $\beta^{2}\sim1.5$} for all SN rates. In reality, $\beta$ should be
smaller than unity for the models H and F -- but since the quenching
of the wind and pumping terms is not considered in the analytic
approach, we obtain artificially high values from
Eq.~(\ref{eq:b_evol}), again illustrating the importance of including
proper vertical transport processes in the dynamo model.


\section{Summary of results}
\label{sec:summary}

The main results of the work that we have presented in the previous
sections can be summarised as follows:

\begin{enumerate}
\itemsep6pt

\item We observe a steady exponential amplification of the magnetic
  energy, $E_{\rm m}$, with a fast e-folding time $\sim 100\Myr$
  during the entire kinematic growth phase.

\item In accordance with the derived dynamo numbers, the back-reaction
  of the mean magnetic field onto the turbulence does \emph{not}
  saturate the dynamo, but instead only leads to an increase of the
  growth time.  Dynamo saturation seems only possible via field
  removal by the wind.

\item Large-scale vertical structures (with `S' mode parity) of the
  average magnetic field evolve within a $\Gyr$, with midplane field
  strengths of about $2 - 3\muG$, as seen in
  Fig.~\ref{fig:mean_fields}. Typical scale heights of the vertical
  exponential profile of the averaged magnetic field are about
  $850\pc$ in the central disc ($h_{0}$) and a few $\rm kpc$
  within the upper halo ($h_{1}$) (see Table~\ref{tab:b_scales}).

\item Scale heights of the inner disc $h_{0}$ are similar to the
  observations of nearby spiral galaxies by
  \citet{2011arXiv1111.7081K}, values of $h_{1}$ however are almost
  half as compared to the observations under the assumption of
  equipartition between cosmic rays and magnetic field. The scale
  heights, $h_{i}$, are moreover found to scale roughly as
  $h_{0}\sim\sigma^{0.4}$ and $h_{1}\sim\sigma^{0.8}$, respectively.

\item We do not observe a noticeable contribution of the MRI in our
  simulations, that is, the final ${\mB} (z)$ profile and the ISM
  composition are similar for models with the same SN rate
  irrespective of the initial vertical flux. This is with the
  exception of the model `QS', in which the initially strong `A' mode
  persists in the dynamical phase, presumably due to the
  flux-preserving periodic boundary conditions.

\item During the kinematic phase, and once the mean field is
  sufficiently established (that is, around $\beta=10^{-2}$), the
  ratio of the mean-to-turbulent magnetic fields is found to scale
  with $\sigma^{-0.3}$ (see Fig.~\ref{fig:Bm_Brms}). This agrees with
  our previous results in the kinematic phase
  \citep{2008A&A...486L..35G}, and matches well the observations by
  \citet{2008A&A...482..755C}. This scaling relation does however not
  remain valid during the dynamical phase, which may be related to the
  correlation between midplane density and the SN rate.  It has to be
  tested with a realistic density correlated variation of the SN-rate.

\item The average vertical density profiles to a first order
  scale exponentially with height, and can be approximated with
  typically three individual scale heights (listed in
  Table~\ref{tab:rho_scales}). The widths of the two thick components,
  $r_{1}$, and $r_{2}$, are found to scale roughly as
  $\sim\sigma^{0.4}$.

\item The average Alfv\'en velocities, $v_{\rm A}$, in the outer halo
  are approximately $30\kms$, irrespective of $\sigma$. This is
  roughly equal to the turbulent velocities, $u'$, for the cases with
  lower SN rates. We observe no statistical correlation between the
  Alfv\'en velocities and the mass density, $\rho$, which implies that
  it is possible to define the vertical profile of $v_{\rm A}$ as a
  ratio of averaged profiles of magnetic field and $\rho^{0.5}$.

\item The vertical structure of the ISM is maintained via a total
  pressure ($P_{\rm tot}$) equilibrium in all thermal components (except
  the hot ISM component), this composition remains unchanged during
  the dynamical phase (cf. Table~\ref{tab:temp_comp}).

\item Within the inner thin disc ($|z|<0.5\kpc$), the magnetic energy
  in the dynamical phase is not distributed uniformly in all ISM
  components (cf. Table~\ref{tab:energy_content}). The comparatively
  largest ratio of $E_{\rm m}$ and $E_{\rm kin}$ is associated with the 
  warm ionised medium and lowest with the hot component.

\item The comparatively largest fractions of magnetic to turbulent
  kinetic energy are reached in the models with smaller SN rates. If
  we consider the entire box ($|z| < 2.1\kpc$), this ratio is still
  smaller than unity (i.e., $E_{\rm kin} > E_{\rm m}$), but within the inner disc ($|z|<0.5\kpc$)
  it exceeds 1 for model Q (see Table~\ref{tab:energy_content}).

\end{enumerate}


\section{Conclusions}
\label{sec:concl}

We have presented new results on the amplification of magnetic fields
in the turbulent multi-phase interstellar medium. Extending our
previous work, we have focused on the late evolution phase, where
magnetic fields become dynamically important. Our main motivation has
been to understand the different saturation behaviour at low and high
supernova rates.

We have demonstrated that magnetic fields can be amplified up to
equipartition with the turbulent kinetic energy in direct simulations
of the turbulent ISM.  However, in model F (with a set of parameters
consistent with the Milky Way), the magnetic energy reached only one
fifth of the kinetic energy.  The reason may be an artificial mass
loss by the wind in a still a too small box setup, but generally the
behaviour seen in DNS can be accounted for by simplified mean-field
models. The correlation of the mean azimuthal field with the 
pitch angle found in a sample of galaxies by \citet{2015ApJ...799...35V} 
is consistent with a partial saturation by the wind preserving the 
pitch angle.

As the key result of this paper, we were able to reproduce the
evolution of $\bar{B}_x(z)$ and $\bar{B}_y(z)$ profiles into the
saturated regime using a simple one-dimensional $\alpha\Omega$~dynamo
model with algebraically quenched coefficients -- notably including
the mean vertical flow velocity $\bar{u}_z(z)$. We remark that this
simplified approach should additionally be complemented with
constraints arising from conservation of magnetic helicity
\citep{2007MNRAS.377..874S}, which may become more severe at higher
magnetic Reynolds numbers \citep{2013MNRAS.429.1686D}.

Furthermore, in a more realistic future scenario, one should include
the coexistence of regions with strong and low star formation -- as
this would naturally be the case when looking at galaxies as a whole.
Observable consequences of this situation can however only be studied
in global models.  Because the computational effort of a global
simulation with the resolution similar to the local models still
exceeds current high performance compute facilities, the use of
dynamical mean-field models is the way to further analyse these
questions. The striking agreement seen in the lower right panel of
Fig.~\ref{fig:mean_fields}, however, strongly advertises the
mean-field approach as a powerful tool in understanding the evolution
of the large-scale ordered fields in the diffuse interstellar medium
-- even into the realm of dynamically important magnetic fields.

A further unknown aspect is the influence of cosmic rays on the
amplification and saturation of the dynamo.  The turbulence may
destroy the anisotropic cosmic ray diffusion and therefore inhibit
dynamo action by cosmic rays.  On the other hand, the cosmic ray
pressure may influence the turbulent transport processes and
eventually change the dynamo properties.

\begin{acknowledgements}
This work used the NIRVANA code version 3.3 developed by Udo Ziegler 
at the Leibniz-Institut f\"ur Astrophysik Potsdam (AIP).
The project is part of DFG research unit 1254. 
\end{acknowledgements}

\bibliographystyle{an}
\bibliography{paper}

\end{document}